%% file: main.tex
 \documentclass[conference,compsoc]{IEEEtran}
\ifCLASSINFOpdf
\else
\fi

\usepackage{listings}
\usepackage{amsfonts}
\usepackage{float}
\usepackage{amssymb}
\usepackage{amscd}
\usepackage{amsmath}
\usepackage{microtype}
\usepackage{array}
\usepackage{graphics}
\usepackage{multirow}
\usepackage{graphicx}
\usepackage[font=small,labelfont=bf]{caption}

\usepackage{pifont}
\usepackage{listings}
\usepackage{xcolor}
\usepackage{subfigure}
\usepackage{booktabs}
\usepackage{balance}
\usepackage{threeparttable}
\usepackage[normalem]{ulem}
\usepackage{enumitem,amssymb}
\newlist{checkbox}{itemize}{1}
\setlist[checkbox]{label=$\square$,topsep=2pt,itemsep=1pt,parsep=1pt}
\usepackage{color}
\definecolor{gray}{rgb}{0.4,0.4,0.4}
\definecolor{darkblue}{rgb}{0.0,0.0,0.6}
\definecolor{cyan}{rgb}{0.0,0.6,0.6}

\usepackage[obeyspaces]{url}

\usepackage[
  breaklinks = true,
  unicode = true,
  urlcolor = blue,
  colorlinks = true,
  citecolor = blue,
  linkcolor = blue]{hyperref}

\usepackage{xcolor, soul}

\newcommand{\hlt}[1]{{#1}}
\newcommand{\hlta}[1]{{#1}}

\newcommand{\ignore}[1]{}
\newcommand{\revised}[1]{}
\newcommand\comment[1]{}

\newcolumntype{L}[1]{>{\raggedright\let\newline\\\arraybackslash\hspace{0pt}}m{#1}}
\newcolumntype{C}[1]{>{\centering\let\newline\\\arraybackslash\hspace{0pt}}m{#1}}
\newcolumntype{R}[1]{>{\raggedleft\let\newline\\\arraybackslash\hspace{0pt}}m{#1}}

\usepackage{pifont}

\usepackage{tikz}
\newcommand*\circled[1]{\tikz[baseline=(char.base)]{
            \node[shape=circle,draw,inner sep=0.7pt] (char) {#1};}}

\usepackage{algorithm}
\usepackage{algorithmic}
\usepackage[misc]{ifsym}
\usepackage{bbding}

\begin{document}

\title{An Internet-wide Penetration Study on NAT Boxes via TCP/IP Side Channel}

\author{\IEEEauthorblockN{Xuan Feng}
\IEEEauthorblockA{Microsoft Research \\ Xuan.Feng@microsoft.com}
\and
\IEEEauthorblockN{Shuo Chen}
\IEEEauthorblockA{Microsoft Research \\ Shuo.Chen@microsoft.com}
\and
\IEEEauthorblockN{Haining Wang}\IEEEauthorblockA{ Virginia Tech \\ hnw@vt.edu}

}

\maketitle

\begin{abstract}

Network Address Translation (NAT) plays an essential role in shielding devices inside an internal local area network from direct malicious accesses from the public Internet.
However, recent studies show the possibilities of penetrating NAT boxes in some specific circumstances. 
The penetrated NAT box can be exploited by attackers as a pivot to abuse the otherwise inaccessible internal network resources, leading to serious security consequences.
In this paper, we aim to conduct an Internet-wide penetration testing on NAT boxes. The main difference between our study and the previous ones is that ours is based on the TCP/IP side channels.
We explore the TCP/IP side channels in the research literature, and find that the shared-IPID side channel is the most suitable for NAT-penetration testing, as it satisfies the three requirements of our study: generality, ethics, and robustness.
Based on this side channel, we develop an adaptive scanner that can accomplish the Internet-wide scanning in 5 days in a very non-aggressive manner.
The evaluation shows that our scanner is effective in both the controlled network and the real network.
Our measurement results reveal that more than 30,000 network middleboxes are potentially vulnerable to NAT penetration. They are distributed
across 154 countries and 4,146 different organizations, showing that NAT-penetration poses a serious security threat. Our work
sounds an alarm for the security community to be aware and tackle
such an Internet-wide security threat caused by legacy problems.

\end{abstract}

%
\IEEEpeerreviewmaketitle

\input{1_intro}

\input{2_prestudy}


\input{3_approach}

\input{4_scanner}

\input{5_measurement}

\input{6_discussion}

\input{7_relatedwork}

\input{8_conclusion}






%



{\footnotesize \bibliographystyle{acm}
\bibliography{ref}}


\end{document}

%% file: 1_intro.tex
\section{Introduction}
\label{sec:introduction}
As a middlebox deployed in network gateways for translating an IP address space into another, 
network address translation (NAT) has played an important role
in network perimeter security for traffic monitoring and filtering.
A NAT box allows IP packets to travel between the internal local area network (LAN) and the external wide area network (WAN), but blocks unsolicited inbound IP packets from the WAN. 
By design, a NAT box is a one-way valve to guarantee that the local network devices with private IP addresses are not routable and thus invisible from the public Internet. 
From a security perspective, NAT is essential to shielding these local devices from direct malicious accesses.

However, recent studies~\cite{mi2019resident,rytilahti2020using} suggest that some application-layer middlebox protocols can be exploited by attackers from outside to penetrate NAT boxes.
Attackers leverage the penetrated NAT box as a pivot to abuse the otherwise inaccessible internal network resources. 
Thus,
such a penetration of NAT boxes 
can lead to serious security breaches. 
Vulnerable local devices can be comprised or become a victim of 
 spoofed-source attacks (e.g., DoS attacks).
Imagine the Internet-of-Things (IoT) scenarios such as smart homes, industrial controls, and medical systems. Launching a simple DoS attack to local IoT devices behind NAT from an arbitrary Internet source can pose serious physical-safety threats to human users.

\vspace{6pt}\noindent\textbf{Motivation.}
{\color{black} Although the previous studies show NAT-box penetrations through application-layer protocols~\cite{rytilahti2020using} 
or specific application vulnerabilities~\cite{acar2018web}, 
we suspect that a more general threat may exist at the TCP/IP layer.
Because a NAT-box by design prohibits an external attacker from connecting to an internal host, we believe that the only possibility for TCP/IP-layer penetration is to exploit blind off-path TCP/IP side channels, similar to ~\cite{cao2016off,feng2020off,alexander2015off,chen2018off}.
Accordingly, a focus of our study is to understand and explore TCP/IP side channels. If such a side channel can be exploited for penetration, we call it a \textit{NAT-penetration side channel}. It 
enables an attacker to identify the IP addresses of the hosts behind a NAT box, and purposefully send packets to reach them.
}

We emphasize three requirements~\label{subsec:goals} for the underlying technique in our study: \textit{Generality}, \textit{Ethics}, and \textit{Robustness}.  TCP/IP side channels have higher generality than the application-level weaknesses exploited by the existing methods. Ethics is important because we are white-hat researchers. The purpose of a penetration test is to assess the security strength of an IT infrastructure by non-aggressively probing its weaknesses.
Our study 
should not disrupt ISPs and their customers. Robustness is also important because observations based on side channels are often affected by noise, as they are indirect. Our approach needs to be sufficiently accurate despite the noise.

\vspace{6pt}\noindent\textbf{Approach.} The first part of our work is to study and compare TCP/IP side channels documented in the research literature (Section~\ref{subsec:sc_investigation}). We find that three of them can be leveraged as the NAT-penetration side channels -- TCP backlog, TCP challenge ACK, and shared IPID counter. However, the former two do not satisfy our requirements (TCP backlog does not satisfy ethics, and TCP challenge ACK does not satisfy ethics and generality).
Only the shared IPID counter~\cite{ipid-original} can be easily inferred without filling the TCP backlog or reaching the rate limit, making it feasible to conduct an ethical and continuous Internet-wide penetration study.
We confirm that there are more than 20 million hosts around the Internet having a shared IPID counter, so the generality of this side channel is the best among all the side channels known to the research community.

The second part of our work is to scan the Internet for discovering penetration holes. The basic idea of our technique is briefly described here. To test every local network, we need to find a publicly accessible host in it with a shared IPID counter, and induce the host to send IP packets inside the local network, thus probe the existence of NAT penetration.
The public host is a passive responder. How to induce it to send packets to private  addresses? The only possibility is that our scanner sends its packets with the source addresses spoofed as the private addresses, so that when the public host ``responds'' to our scanner, it actually sends packets to the private addresses. In this paper, we call such a public host, which responds to the source-spoofed packets, an \textit{outpost} of the local network. The outposts act as the helpers in the scanning. On the other hand, since the ``responses'' from the outposts are not sent to our scanner, we need to utilize the side channel information (i.e., IPID) to infer whether the NAT penetration indeed happens.

\vspace{6pt}\noindent\textbf{Challenges.} The idea is conceptually simple, but deploying the method in the Internet-wide scale faces challenges. On one hand, the absence of other connections and the packet losses on an outpost will induce unexpected IPID increments. 
It is hard to determine whether an IPID increment is triggered by our method or other communications (i.e., noise for us).  On the other hand, there are a large number of targets (about 40 billion IP addresses) around the world. The Internet is complex and heterogeneous, making an adaptive and robust scanning tool difficult to build.

Facing the challenges, in the second part of this work, we develop a scanner, which can automatically adjust its strategy and make robust decisions for different network scenarios. 
The scanner sends IP packets at a very low rate, 0.6 packets per second on average. Despite such a low rate, it accomplishes the IPv4-scale scanning in 5 days.  
Basically the scanning approach consists of three steps. In the first step, we target the entire IPv4 address space to identify all hosts that are alive and potentially vulnerable to TCP/IP side channels. Our technique only needs to send two packets to test every IP address.  
In the second step, we analyze the IPID behavior and IP spoofing capability of every potential host to determine whether it can be used as an outpost. In the third step, we leverage the outpost of each local network to send IP packets to the local addresses and use the IPID side channel to reveal whether the NAT penetration happens.

{\color{black}

Whereas TCP/IP side channels (e.g., shared IPID), ISP security weaknesses (e.g., insufficient ingress filtering against IP spoofing capability), and other security-related issues have been known individually for a long time, the interactions of these issues and their security implications have never been investigated before. In this work, we reveal that the interactions of these legacy security problems can be exploited to pose a serious security risk, middlebox penetration. Specifically, we conducted a large scale security measurement study of NAT-penetration on the Internet and show our new findings.
}

\vspace{6pt}\noindent\textbf{Contributions}. 
The major findings of our study are summarized as follows:

\vspace{3pt}\noindent$\bullet$ 
We identify more than 30,000 outposts.
The local networks behind them are potentially vulnerable to NAT penetration.
The outposts have a significant presence across 154 countries and 4,146 different organizations. The vulnerable outposts and local networks are located in both big ISPs (like TalkTalk and LG DACOM Corporations) and small ISPs. 

\vspace{3pt}\noindent$\bullet$ 
We identify the types of these vulnerable outposts. They range from routing switching devices, to WINDOWS hosts, to different kinds of IoT devices. Some edge routing switching devices from well-known manufacturers are among the vulnerable outposts.

\vspace{3pt}\noindent$\bullet$  To provide a convincing evidence on the serious consequence of NAT penetration, we reconfirm the feasibility of a targeted IoT-device DoS attack (i.e., resetting a TCP connection on a NAT-protected device) in our own test-bed. This end-to-end demonstration shows concretely that addressing the NAT penetration problem by the Internet community is truly needed.  

\vspace{3pt}\noindent$\bullet$ The root causes behind the NAT penetration problem are the shared-IPID vulnerability and the misconfigurations that fail to block spoofed inbound private-IP addresses. Our investigation shows that the legacy IPID vulnerability, known to the public since 1998, is still pervasive on the Internet. 
{\color{black}
Regarding the inbound filtering of spoofed packets, 
56.2\% of the hosts with the ideal global IPID fail to enable an appropriate ingress filtering.
}
This problem is found in well-known ISPs and device manufacturers, indicating the insufficiency of security incentive/awareness at the network edges.

\vspace{8pt}\noindent\textbf{Roadmap.} The rest of the paper is organized as follows. Section~\ref{sec:overview} gives an overview of our study. 
Section~\ref{sec:scanner}  details on how to conduct an Internet-wide NAT penetration test. 
Section~\ref{sec:impl} describes the implementation of our Internet-wide scanner.
Sections~\ref{sec:measurement} presents the evaluation of revealing the existence of NAT-penetration holes via side channels, and reports our Internet-wide scanning results of the NAT-penetration test.
Section~\ref{sec:discussion} discusses the limitations of our experiments, mitigation actions, and the ethics. 
Section~\ref{sec:relatedwork} surveys related work, and finally, Section\ref{sec:conclusion} concludes the paper.

%% file: 2_prestudy.tex
\section{Overview}\label{sec:overview}

\subsection{Problem Definition}\label{subsec:attack-model}

\begin{figure}[t]
\centering
\includegraphics[width = 0.98\linewidth]{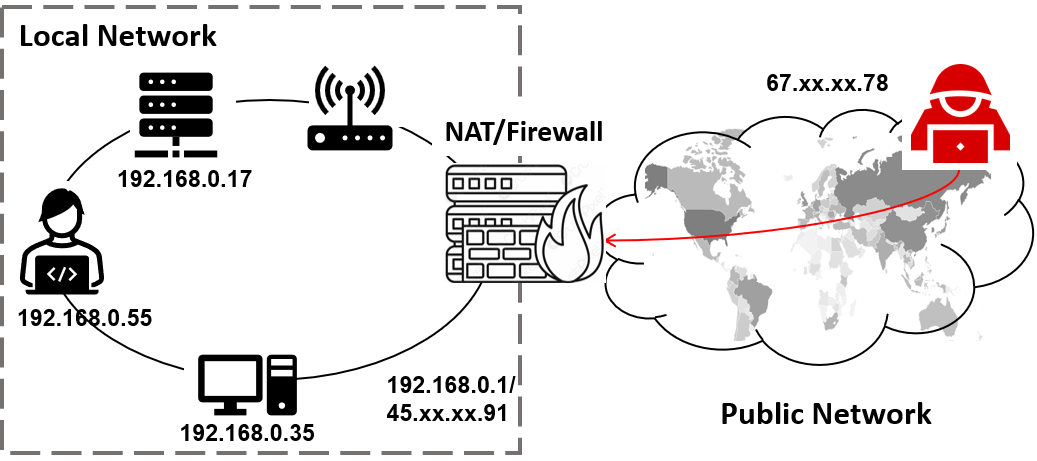}
\caption{The attack model of our study.}
\label{fig:attack-model}
\vspace{-5pt}
\end{figure}

Our study considers the threat scenario shown in Figure~\ref{fig:attack-model}. The attacker (e.g., with the IP address 67.xx.xx.78) is on the public network. A number of network devices with private IP addresses are in a local network (192.168.0.0/24), which connect to the Internet via a public gateway (45.xx.xx.91). 
The gateway disallows any host in the public network to proactively start a connection with a local device. 
However, a host in the public network, potentially the attacker, can exchange packets with the gateway.
Through side channels, the attacker may infer whether they can pivot to the otherwise inaccessible internal networks (e.g., the external attacker can send spoofed packets to 192.168.0.17 penetrating the NAT boxes).
As mentioned before, we call this gateway an \textit{outpost}\footnote{The outpost is an Internet host with public IP address, which can be used to reveal the NAT penetration problem.}, and each of the side channels a \textit{NAT-penetration side channel}. 

By conducting an Internet-wide scanning at the TCP/IP layer for penetration test, we will show the pervasiveness and seriousness of the NAT penetration problem, as well as the types of the NAT-protected devices that can be victimized by attackers.

\vspace{-5pt}

\subsection{Investigating the TCP/IP Side Channels}\label{subsec:sc_investigation}

The generality requirement of our study means that we should not rely on any specific features or behaviors at the application level. Thus, we begin with a study on TCP/IP side channels documented in the research literature. We find three side channels potentially suitable as NAT-penetration side channels -- the \textit{TCP backlog}, the \textit{TCP challenge ACK}, and the \textit{shared IPID counter}.

\vspace{3pt}\noindent\textbf{TCP backlog}. 
TCP connections that have not completed a ``three-way handshake'' are stored in a buffer called the \textit{backlog}, which is shared across all TCP connections waiting either for an ACK to complete the normal handshake process, or an RST, an ICMP error, or an ARP timeout to drop a connection.

For Linux kernel before version 2.3, there is one backlog for holding all uncompleted TCP connections. When the backlog is not full, a regular SYN-ACK packet is sent as a response.
When the backlog is full, the response is a SYN cookie.
The backlog gives a side channel that allows the attacker to determine whether 
a target host is reachable without directly sending any packet to it, as described below~\cite{ensafi2010idle}. 
This attack involves a third machine called a ``zombie'' host.
The attacker sends spoofed SYN packets to the zombie host trying to fill its backlog.
The spoofed SYN packets will trigger the zombie host to respond a SYN-ACK packet to the target. If the port is closed, the spoofed packets will be dropped from the backlog by RST packets. Otherwise, the spoofed packets are still stored in the backlog until it is full, in which case, if the attacker sends a SYN packet (non-spoofing) to the zombie host, a SYN cookie is sent back to the attacker. This reveals that the target host is reachable and the port on the target host is open.
A drawback of this backlog side channel is the possibility of DoS attack due to the filled-up backlog.

For Linux kernel since version 2.3, the backlog entries are categorized as either ``young'' or ``mature''. ``Young'' requests are those that have not been retransmitted yet. 
If the backlog (~\textit{tcp\_max\_syn\_backlog}) is more than half full, some of the older (i.e., ``mature'') entries in the backlog are evicted to make room for ``young'' requests. 
The newer TCP backlog mechanism also allows the attacker to infer whether a target host is reachable without directly sending any packet to it~\cite{zhang2015original}. The attacker uses the similar method as the old TCP backlog does, except that the new method fills the TCP backlog half full to see whether some entries in the backlog are evicted.
However, the process of inferring the TCP backlog size is not trivial which needs to repeatedly guess from the typical range 16 to 256 until the correct one is found. 
In addition, it is required that packets are sent at the rate of 5 packets per second for about 60s, so that the TCP backlog becomes half full. Obviously, this method is too aggressive to be used in a white-hat research study.

\vspace{3pt}\noindent\textbf{TCP challenge ACK}.
RFC 5961 introduces the challenge ACK response~\cite{ramaiah2010rfc} on an established connection. If a guessed sequence number of the spoofed packet happens to fall in the receive window (called an in-window sequence number), the spoofed packet will trigger a TCP challenge ACK. Otherwise, the spoofed packet will be dropped.
In order to reduce the number of challenge ACK packets that waste the CPU and bandwidth resources, a system-wide rate limit for the challenge ACK packets (challenge ACK counter) is set.
Specifically, in version 3.6 of Linux, a global variable \textit{tcp\_challenge\_ack\_limit} is introduced to control the maximum number of challenge ACKs generated per second. It is set to 100 by default in Ubuntu 14.04.

This approach, unfortunately, creates an undesirable side channel.
It allows an attacker to create contention on this system-wide rate limit counter on the target system by sending spoofed packets. 
The attacker can then observe the effect on the counter changes, measurable through probing packets.
This feature can be used to infer the presence or the absence of an ongoing connection between a client and a server without targeting the client~\cite{cao2016off}.

In practice, however, the attacker needs to send at least 100 non-spoofed in-window RST packets to exhaust the challenge ACK count. In some newer patched Linux versions, the number is set to 1,000, which is too intrusive for a white-hat study.
Moreover, this method only works when there is an ongoing connection between the client and the server, which does not satisfy our goal of generality.

\begin{figure*}
\centering
\includegraphics[width = 0.66\linewidth]{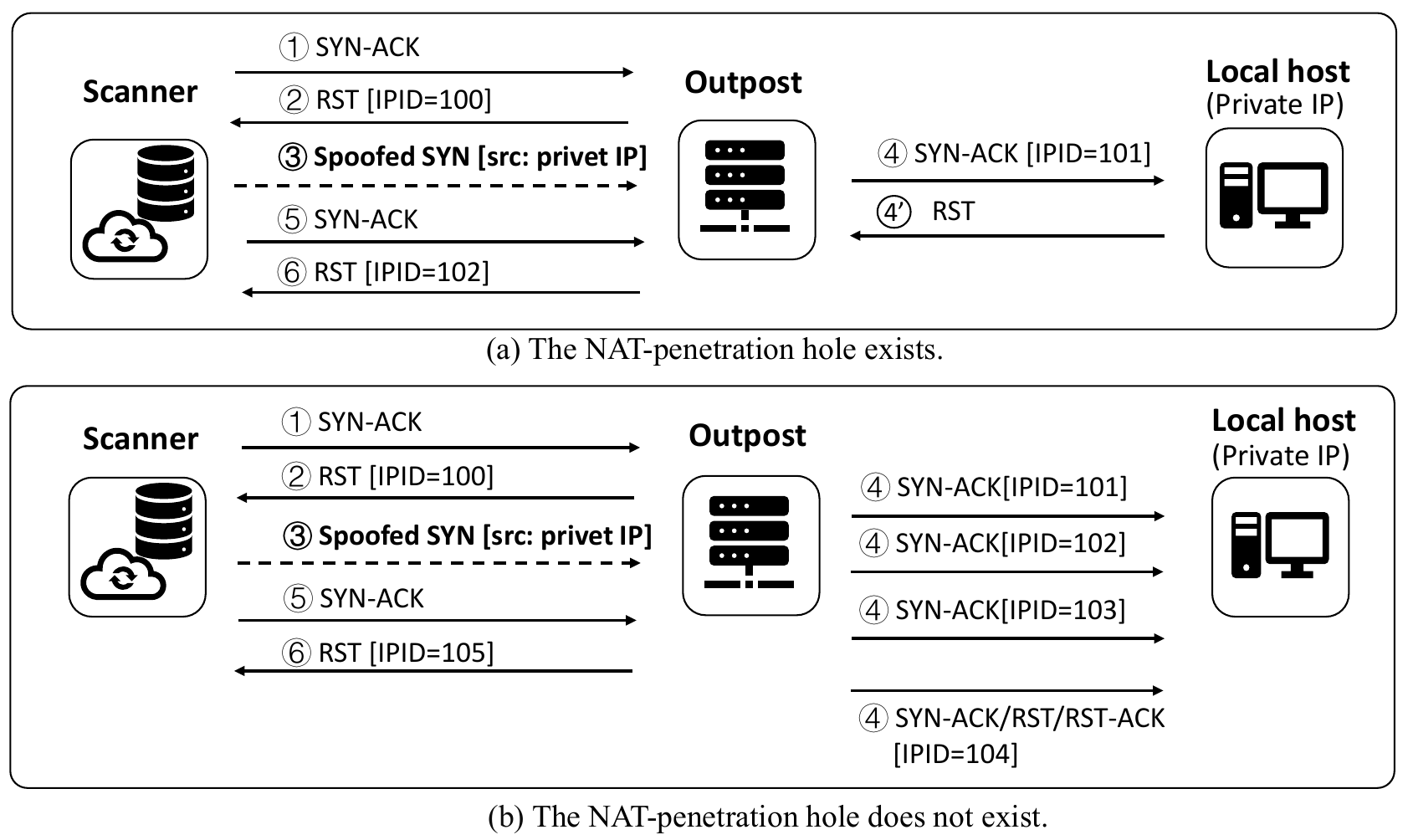}

\caption{Revealing the NAT-penetration hole. The outpost is connected to the Internet with a public IP address and satisfies our requirements. The local hosts are machines that hide behind NAT with private IP addresses. The source IP address of the SYN packet is spoofed as the private IP address of the local host.}
\label{fig:overview}
\vspace{-5pt}
\end{figure*}

\vspace{3pt}\noindent\textbf{Shared IPID counter}.\label{subsec:ipid}
Each IPv4 packet contains a 16-bit identification field (or ``IPID''). The purpose is to help endpoints re-assemble fragmented IPv4 packets.

A variety of mechanisms can be used to produce an IPID for each packet, e.g., a random number, a constant number or
a counter increment-by-one per-host/per-interface (i.e., local counter) it communicates with.
However, many hosts use a single {\em global} counter (rather than per-host counter) to increase the IPID value for all packets originated from the host.
In this situation, the IPID value at a given time reflects how many packets the host has generated. Thus, the attacker, who has the ability to observe the IPID value over time, can infer whether a host is generating IP packets and how many.

Similarly, this side-channel allows the attacker to identify whether a port is open on a target host without directly sending any packets to the target~\cite{ipid-original},
or determine whether the connection between two Internet locations is reachable without directly controlling a measurement vantage point at either location~\cite{ensafi2014detecting}.
Compared to filling the TCP backlog or reaching the rate limit, this side-channel is much easier to exploit, because it essentially requires only 3 packets.

The shared-IPID side channel, dated back to 1998~\cite{ipid-original}, is an old problem. Some researchers have a false perception that 
the number of Internet hosts with the shared IPID counter is rare~\cite{zhang2015original}, because modern Windows, Linux, Mac, and BSDs do not use shared IPID anymore. 
However, as shown in the recent research~\cite{salutari2018closer}, although the constant IPID number (34\%) and the local counter (39\%) are now prevalent, the single {\em global} counter implementation still occupies a considerable proportion (18\%). 
Considering the whole IPv4 address space, the absolute number is massive. The prior research~\cite{pearce2017augur} reported that the number is approximately 22.7 million (slightly more than 16\% among 140 million reachable hosts). 
These hosts are geographically distributed across 234 countries. We re-validate these numbers in our study (see Section~\ref{subsec:world-setup}).
We believe that the shared IPID is the most general and lightweight side channel known to the research community, and thus it is selected for our Internet-wide NAT penetration study.

%% file: 3_approach.tex
\subsection{Revealing the NAT-penetration Hole}\label{subsec:method}

Up to now, TCP/IP side channels  have only been used in various operations targeting public hosts on the Internet, but they have never been used to infer the reachability of the local networks behind NAT.
The core of our work is to find a feasible technique to use the shared-IPID side channels to perform the NAT penetration testing. Our approach is briefly summarized here. 
In Section~\ref{sec:scanner}, we provide the details on how to conduct the Internet-wide NAT penetration study.

\vspace{3pt}\noindent\textbf{Approach}.\label{subsec:approach}
The overview of our NAT penetration testing procedure is as follows. On the Internet, we first find the outposts with the shared IPID counter. For each outpost, we induce it to send packets inside its local network. Through the IPID side channel, we can infer whether the packets indeed reach the end hosts inside the local network (i.e., the existence of the NAT-penetration hole)
without directly controlling a measurement vantage point in the local network.
Specifically, our scanner sends a spoofed packet to the outpost with the source IP address set as the private IP address.
Normally, the outpost responds a packet to the private IP address which triggers further communication between the outpost and the private IP address. 
Because the private communication cannot be observed externally, the scanner uses the shared IPID counter to infer whether the communication (NAT penetration) happens and how many packets the outpost sent.

The procedure is shown in Figure~\ref{fig:overview}.
To monitor the IPID increment of the outpost, the scanner first sends a TCP SYN-ACK packet to the outpost (\ding{172}) and records the IPID value in the responded TCP RST packet (\ding{173}). We choose a SYN-ACK packet, instead of a SYN packet, because the SYN-ACK packet is responded with an RST packet, which quickly terminates the TCP connection without inducing more loads on the target host as the SYN packet did. 
To infer the existence of a NAT-penetration hole, the scanner sends a spoofed TCP SYN packet to the outpost (\ding{174}), the source IP address of the spoofed SYN packet is a private IP address (e.g., 192.168.1.1.). 
If there is no ingress filtering to reject IP-address spoofing, the spoofed SYN packet from the scanner to the outpost triggers a SYN-ACK packet from the outpost to the private IP address (\ding{175}).

If the NAT-penetration hole exists on the NAT box, the private IP address in turn triggers a TCP RST to the outpost in order to end this session (\circled{\scriptsize 4'}), because the private IP address has not previously set up this connection with a TCP SYN packet.
As mentioned earlier, the outpost generates IPID values for the packets based on a single shared counter, the connection between the outpost and the private IP address shares the same IPID counter with the connection between the outpost and the scanner. In an ideal scenario, the scanner can observe how many packets the outpost exchanges with the private IP address.
When we examine the IPID of the outpost again (\ding{176}\ding{177}), the IPID counter increases by two. Figure~\ref{fig:overview}(a) shows this process.

If the NAT-penetration hole does not exist or the private IP address does not exist
(i.e., no response comes back), the outpost continues to retransmit the SYN-ACK packets multiple times (typically between 3 and 5 times) with specific time intervals based on the outpost’s operating system (\ding{175}). This causes further increments in the IPID value of the outpost. Figure~\ref{fig:overview}(b) shows the case where four SYN-ACK packets are retransmitted. We can see that the IPID counter increases by five (\ding{177}). Therefore, the scenarios in Figure~\ref{fig:overview}(a) and Figure~\ref{fig:overview}(b) can be distinguished.

%% file: 4_scanner.tex
\section{Methodology and Techniques}\label{sec:scanner}

The procedure described above assumes an ideal scenario without packet losses or sudden bursts of packets caused by other connections.
However, the Internet is complex and dynamically changing, which poses challenges for the Internet-wide assessment.
The presence of other connections and the packet losses on an outpost cause unexpected IPID increments. It is hard to determine whether an IPID increment is caused by our procedure or other communications. 
Another challenge is the large number of targets (about 40 billion addresses) in the complex and heterogeneous Internet, making an adaptive and robust scanning tool difficult to build.
In this section, we explain in details how we conduct the Internet-wide penetration test despite these challenges.

\begin{figure}
\centering
\includegraphics[width = 0.99\linewidth]{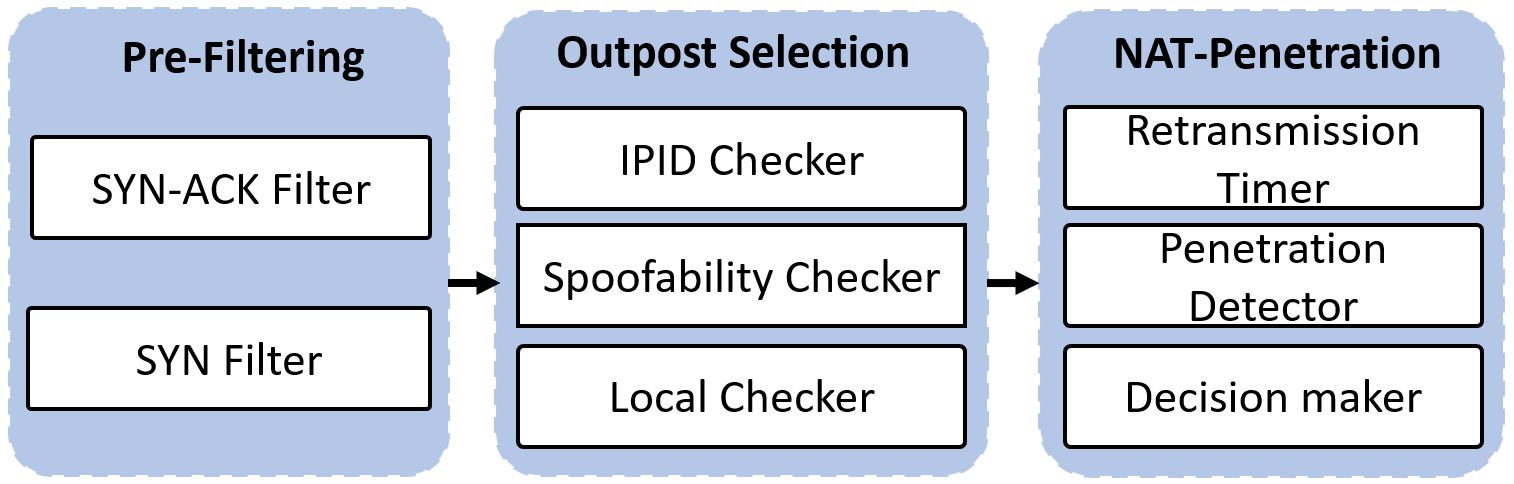}
\vspace{-10pt}
\caption{The workflow of our scanner.}
\label{fig:scanner}
\vspace{-5pt}
\end{figure}

\subsection{Architecture of Scanner}
\label{subsec:outpost_requirement}
To address these challenges, we present an adaptive ethical scanner, which can automatically adjust and make robust decisions in different networks. The scanner can send packets at a very low rate (0.6 packet per second) and complete the Internet-wide scanning within 5 days, much more efficient than the previous studies.

Figure~\ref{fig:scanner} illustrates the architecture of the scanner, which has three major components: (1) pre-filtering, (2) outpost selection, and (3) NAT-penetration detection. 
The pre-filtering is used to quickly test the whole IPv4 address space to identify all hosts that are alive and potentially vulnerable to TCP/IP side channels. 
After the filtering, the outpost selection is used to check the properties that the outposts should satisfy.
For each selected outpost, we use the NAT-penetration detection to determine whether a penetration hole indeed exists inside its local network.

Before explaining the details of our method, we elaborate on the requirements for outposts. An outpost needs to satisfy the following properties:

\vspace{3pt}\noindent$\bullet$\textit{ Shared IPID counter.}  
As described in Section~\ref{subsec:method}, our method uses the shared IPID counter as the side channel to infer the existence of NAT-penetration holes.
This is the key requirement. A previous work~\cite{pearce2017augur} shows that ``approximately 22.7 million IP addresses demonstrated use of a shared, monotonically increasing IPID'' in the whole IPv4 network.

\vspace{3pt}\noindent$\bullet$\textit{ No ingress filtering.}
Our method relies on the spoofed SYN packets to induce the communication between the outpost and the private IP address. Therefore, one requirement of the target's network is that it does not have ingress filtering against source-IP spoofing.

\vspace{3pt}\noindent$\bullet$\textit{ Normal TCP/IP state machine.} The outpost should have some ports open and conform to the TCP/IP protocol specifications. Specifically, the outpost should respond to a SYN packet with a SYN-ACK packet. If there is no response to the SYN-ACK packet, the outpost should retransmit the SYN-ACK packet multiple times. In addition, if the outpost receives the SYN-ACK packet (without sending the SYN packet before), an RST packet should be generated to reset this session. 
Note that these requirements are easy to satisfy in reality because they comply with the TCP/IP protocol specifications.

\subsection{Pre-Filtering}
The entire IPv4 address space consists of 40 billion addresses. It will be extremely time-consuming to test all the requirements for every single host in the address space.
The necessary first step is to use as few packets as possible to filter out obviously ineligible hosts, because current single-packet network scanners~\cite{masscan,durumeric2013zmap} can conduct the Internet-wide scanning within one hour. 

By analyzing the requirements of the outpost, we find that SYN-ACK probes and SYN probes can quickly test the entire IPv4 address space to identify all hosts that are candidate outposts.
For SYN probes, we will filter out the hosts that do not respond to a SYN-ACK packet,  ensuring that the target host is alive and with a port open. 
The hosts passing the first filtering are tested by SYN-ACK probes,
which aims to filter out the hosts that obviously do not have a shared IPID counter. 
In each SYN-ACK probe, besides checking the responded RST packet, we also filter out every host whose IPID value is equal to zero, because the probability that a shared IPID counter happens to be zero is very low (1/65536 = 0.0015\%).
The hosts passing these two probes will be kept for the next outpost selection stage.

\subsection{Outpost Selection}\label{subsec:outpost_selection}

{\color{black}
After the pre-filtering, we filter out hosts that obviously do not have a shared IPID and cannot satisfy our outpost requirements. Then, we run an outpost selector to test the properties described above.} The selector includes three components: the \textit{IPID checker}, the \textit{spoofability checker}, and the \textit{local checker}.
The IPID checker is used to test whether a host has a shared IPID counter. The spoofability checker and the local checker, based on the shared IPID counter, determine whether the ingress filtering (including the spoofed public and private IP addresses) exists on the target host's network. The hosts passing these three testings can be used as the outposts to reveal the existence of NAT-penetration holes. (Section~\ref{subsec:private_ip_detection}).

\vspace{3pt}\noindent\textbf{The IPID checker}.
To check whether a host has the shared IPID counter, we need to monitor the IPID value on the host over time. 
We continuously send SYN-ACK packets to the target host and record the IPID value of RST packets. 

This monitoring module is critical to the scanner as it is used by different functionalities of the scanner.
For convenience, we call it the \textit{IPID probe module}. 
The IPID probe module sends $N$ SYN-ACK probes to each host with a time interval $T$ seconds between two probes.
In the end, the IPID probe module outputs an array $ipids(T, N) = [ipid_1, ipid_2, ipid_3, ..., ipid_n]$.
To further ensure that this side channel is stable and reliable enough
to support continuous measurements, the probe module also records a `None' flag if it does not receive an RST packet before the TCP connection timeout (1 second in our work). For example, if the second packet is timeout, the $ipid_2$ is recorded as `None'.
The number of `None' entries in $ipids(T, N)$ indicates the network quality on the target host. 
Because the IPID probe module causes the packet generation on the outpost at a constant rate (one per second), any additional IPID increments must be due to the traffic from other connections. The additional IPID increments acquired by the IPID probe module gives us the opportunity to estimate how many other connections (i.e., how much IPID-interference) exist on the target host. This provides the guidance on how many spoofed packets we should send to dominate the IPID interference.

We run the IPID probe module for each candidate outpost produced by the pre-filtering. 
This probe module outputs the $ipids0$.
We check the $ipids0$ to see if it satisfies the following two conditions: 

\vspace{3pt}\noindent$\bullet$\textit{ The number of `None' in $ipids0$}. We remove the hosts having too many `None' entries, indicating the poor quality of the current network condition. The removal ensures that each outpost is stable and reliable enough to support continuous measurements.

\vspace{3pt}\noindent$\bullet$\textit{ The $ipids0$ is monotonically increased}. The host with the shared IPID counter exhibits a monotonic upward trend in the IPID value.
The increment of the IPID value, which the scanner can observe, implies the communication between the outpost and other Internet endpoints.

\vspace{3pt}\noindent\textbf{The spoofability checker}.\label{subsec:spoofing-checker}
The ingress filtering~\cite{bcp38} is a firewall setting that is used as an effective countermeasure against various spoofing attacks. This setting is a black box for our scanner.
To test its existence, one possible way is to send some spoofed packets and check whether the target machine receives spoofed packets and give  responses. 
Since the ``receive-and-respond'' status cannot be observed by our scanner, we need to  utilize the IPID side channel to infer this status. 
Specifically, we set up two IPID probe modules before ($ipids1$) and after ($ipids2$) a spoofed packet is sent. 
If the target host receives the spoofed packet and responds to it, the IPID value that the host generates will increase by 1. We can observe the increment by comparing $ipids1$ and $ipids2$.

However, as mentioned earlier, because the target host may have multiple connections in a real network environment, it is hard to determine whether the IPID increment is induced by our method or other communications, which is \emph{noise} for us. 
To increase the ``signal-to-noise'' ratio, we must increase the number of spoofed SYN packets. We send $M$ spoofed SYN packets to the target host. 
The value of $M$ depends on the current network noise level, which is calculated based on the average increment of $ipids1$. For example, if the average increment is 2, $M$ will be set to a number larger than 2, such as 4.

By checking the IPID increment between $ipids1$ and $ipids2$, we determine whether the ingress filtering exits. We have confirmed that there is no egress filtering on our scanner (see Section~\ref{subsec:scanner_requirement}) and the IPID increment induced by ours is larger than the IPID noise. 
If the IPID value is observed to increase by $M+1$, it indicates that the outpost's network does not filter spoofed IP packets.

\vspace{3pt}\noindent\textbf{The local checker}. 
The difference between the local checker and the spoofability checker is
that the source IP address of a spoofed SYN packet changes to a private IP address (e.g., 192.168.1.1). And we start a new IPID probe module ($ipids3$) after the spoofed packet is sent. 
Again, if the IPID has increased by $M+1$, we determine that the target network does not filter spoofed packets with private IP addresses. 
For better robustness, if we find a host not blocking spoofed packets with private IP addresses, we repeat the last two step to confirm.

\begin{figure}
\centering
\includegraphics[width = 0.83\linewidth]{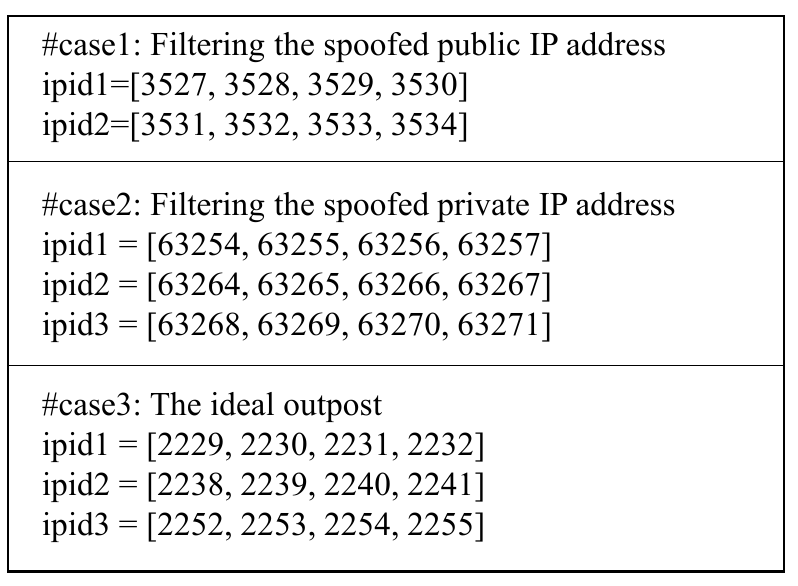}
\vspace{-2pt}
\caption{Examples of the outpost selection.}
\label{fig:example1}
\vspace{-10pt}
\end{figure}

\vspace{3pt}\noindent\textbf{Examples}. Figure~\ref{fig:example1} shows some examples. In these cases, 5 spoofed packets are sent. We expect the IPID of the outpost to increase by 6. 
In the first case, we do not observe an expected IPID increment. This indicates the existence of ingress filtering for both spoofed public IP addresses and spoofed private IP addresses. The second case only has an expected IPID increment between $ipids1$ and $ipids2$, indicating that only the spoofed private IP address is blocked. The third case shows an ideal outpost. It has two IPID increments ($ipids1$ and $ipids2$,  $ipids2$ and $ipids3$).

\subsection{NAT-penetration Detection}~\label{subsec:private_ip_detection}
Section~\ref{subsec:approach} briefly describes the basic idea of our approach to reveal the existence of a NAT-penetration hole based on a selected outpost. If the NAT-penetration hole does not exist, the outpost continues to retransmit SYN-ACK packets for a fixed number of times, resulting in more IPID increments. Hence, we can observe the difference (e.g., the IPID counter increases by five instead of two).
However, in a real-world scenario, one SYN-ACK retransmission may take a long time to complete, which makes the IPID noise more substantial.
{\color{black} Because of the ethics requirement, we cannot simply increase the number of spoofed SYN packets.
The previous study deals with the similar challenge via a statistical method (sequential hypothesis testing)~\cite{pearce2017augur}.  
This statistical method depends on repeated trials and the trials must be independent and identically distributed. To achieve this condition, they run experiments over the course of weeks.
However, the NAT-penetration hole is quite dynamic, which needs to be identified in a timely manner (within a few minutes). 
Obviously, the previous method cannot be directly used in our work.
}

To handle this issue, we introduce a new method based on the timing characteristics of the retransmission mechanism.
In addition to checking the total IPID increment, our method also 
checks whether the IPID increment occurs in the right timing to dominate the noise.
Specifically, we precisely identify the time point when the SYN-ACK packet is retransmitted, and test this point to observe whether the IPID increase indeed happens.
For example, the outpost retransmits SYN-ACK packets 1/3/7/15 seconds after the first SYN-ACK packet is responded. In addition to checking that the IPID is incremented by 5, we also check whether each IPID increment occurs at the related time points (e.g., 1/3/7/15).

\begin{figure}
\centering
\includegraphics[width = 0.85\linewidth]{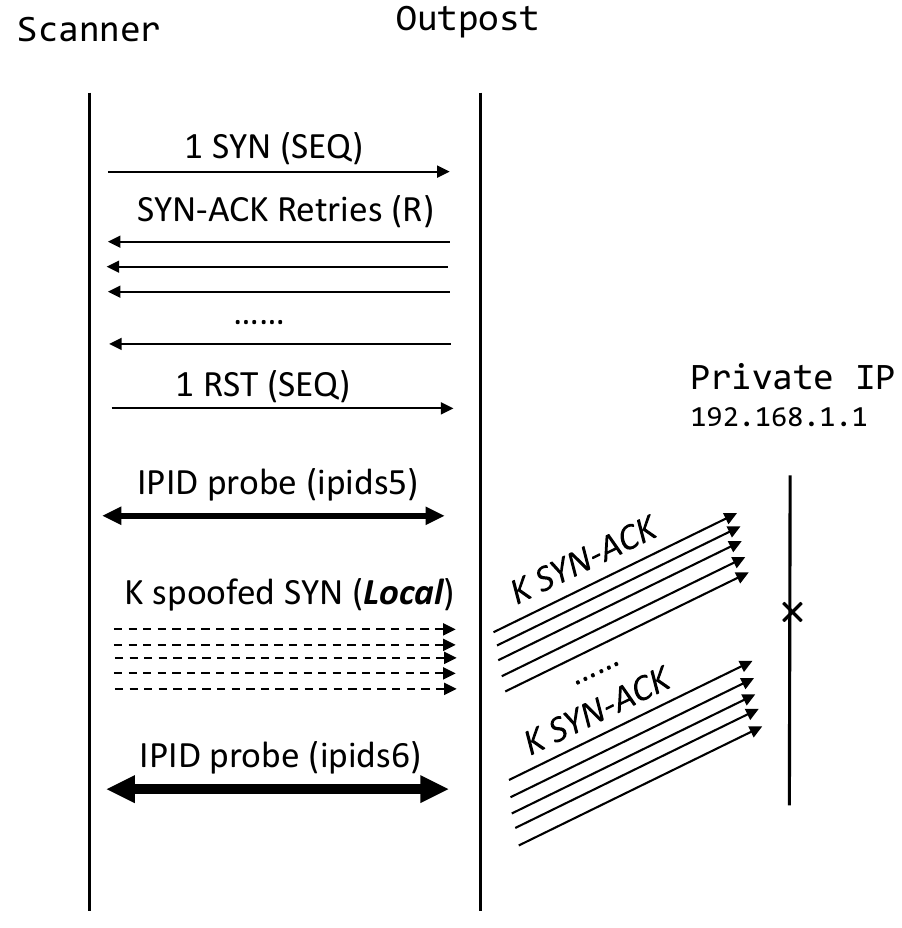}
\vspace{-5pt}
\caption{Revealing the NAT-penetration hole.}
\label{fig:local}
\vspace{-10pt}
\end{figure}

The complete steps of our method are shown in Figure~\ref{fig:local}. 
First, we propose a retransmission timer to identify the retransmission details for each outpost. 
Then, we send spoofed SYN packets to the outpost trying to penetrate the NAT boxes and record the outpost's IPID increment. 
Finally, the decision maker compares the retransmission details and the IPID increment to make a decision whether the NAT-penetration hole exists.

\vspace{3pt}\noindent\textbf{Retransmission timer}. 
When and how many SYN-ACK packets should be retransmitted differ among the outposts, because they depend on the settings of the operating systems. 
Fortunately, for one specific outpost, this setting remains stable for a relatively long time, which gives us the opportunity to test each outpost to know its retransmission details in advance.
Our method works as follows.
We send a SYN packet to the outpost. Then, the outpost responds with a SYN-ACK packet back to our scanner.
Normally, the kernel of our scanner will reply an RST packet to end this session.
If our scanner does not send back any packet, the outpost will retransmit the SYN-ACK packets until the timeout elapses.
To force the outpost to retransmit the SYN-ACK packets, our scanner intentionally drops the responded RST packets from the kernel (by using an \textit{iptables\cite{iptables}} firewall rule). 
A typical TCP/IP stack retransmits a SYN-ACK packet 5 times, the general retransmission interval increases exponentially. For example, if the first interval is 3 seconds, the next interval will be 6 seconds. Assuming the first SYN-ACK packet is sent at time 0, the TCP/IP stack would retransmit the SYN-ACK packets at times -- 1, 3, 7, 15, 31 seconds or 
 3, 9, 21, 45, 93 seconds.
We monitor this retransmission (triggered by a normal SYN packet) session for 25 seconds, 
which covers at least 3 SYN-ACK retransmissions.
Meanwhile, we record the timestamps when the retransmission occurs in an array --- $R$. For example, if the TCP/IP stack retransmits  SYN-ACK packets at times --- 1, 3, 7, 15 seconds, then $R = [1, 3, 7, 15]$.
The $R$ later will be used in the \textit{decision maker} to provide the  
the time point we need to check whether the IPID increment indeed occurs.

A difficulty we face is that the retransmission timer sometimes may shift 1 second to the left or the right due to the network jitters. For example, 
we have $R = [5, 17]$, but In fact R is [6, 18]. 
The wrong R will make us check the wrong time point, leading to the wrong decision.
Therefore, we add a heuristic timer calibration, which calibrates $R$ to the right value (e.g., $[6, 18]$).
Specifically, when we have $R = [5, 17]$, 
we infer that the outpost retransmits SYN-ACK packets at times 5 and 17 seconds.
However, it implies that the first retransmission interval is 5, and the second interval is 12. This is contradictory with the typical TCP/IP stack specification that the next interval should be doubled compared with the current interval. 
We determine that this R must be wrong. Then, we start the timer calibration.
We increase or decrease every number in $R$ (we get $R=[6, 18]$) by 1 and check again. If the updated $R$ satisfies the TCP/IP specification, we use it to replace the original (Finally, $R=[6, 18]$). Otherwise, we keep the original value.

\vspace{3pt}\noindent\textbf{The penetration detector}. 
After the timing, 
we send $K$ spoofed SYN packets to the outpost, the source IP address of each spoofed SYN packet is a private IP address (e.g., 192.168.1.1).
{\color{black}The value of $K$ also depends on the observed noise level, which is calculated based on the average increment of $ipids5$.} 
Because we have ensured that the ingress filtering for the spoofed packets does not exist (Section~\ref{subsec:outpost_selection}). 
The $K$ spoofed SYN packets from the scanner to the outpost elicit $K$ SYN-ACK packets from the outpost to the private IP address. 
The two IPID probe modules are set up before ($ipid5$) and after ($ipid6$) the spoofed packets. 
Note that there is a 0.5-second delay between the last spoofed packet is sent and the new IPID probe module ($ipid6$) starts, which ensures all the IPID increments are observed.
To avoid further IPID increments, we send $K$ following RST packets to end the retransmissions.

\begin{figure}
\centering
\includegraphics[width = 0.9\linewidth]{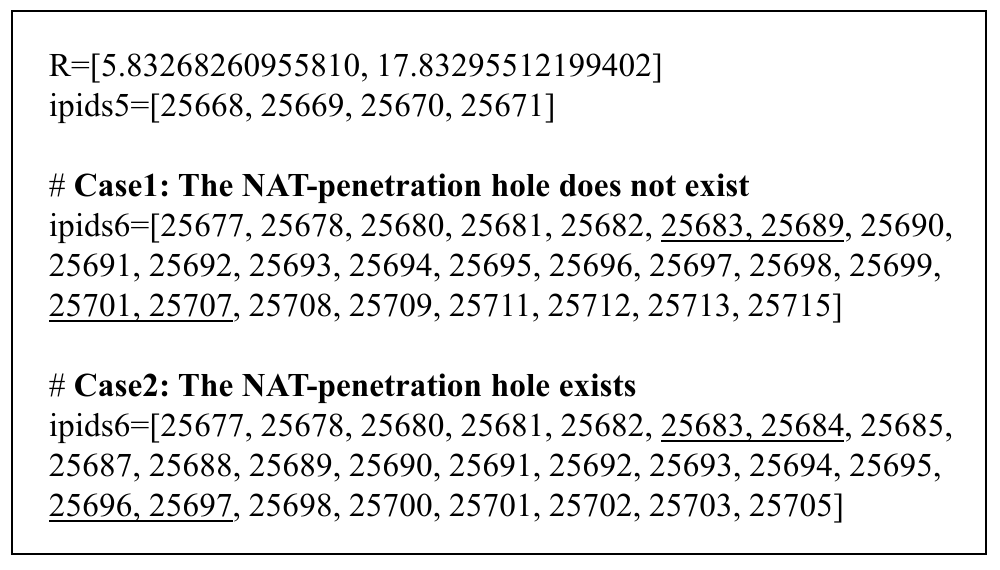}
\vspace{-1pt}
\caption{Examples of the NAT-penetration detection.}
\label{fig:example2}
\vspace{-10pt}
\end{figure}

\vspace{3pt}\noindent\textbf{The decision maker}. 
By checking the time points when the SYN-ACK packets are retransmitted ($R$, we get in retransmission timer) in $ipid6$, 
we can infer whether the NAT-penetration hole exists.
An example is shown in Figure~\ref{fig:example2}. After the calibration, we get $R = [6, 18]$, which indicates that the SYN-ACK packets are retransmitted at sixth and eighteenth seconds in $ipids6$.
We check the IPID increment between $ipids6[6]$ and $ipids6[5]$ and the increment between $ipids6[18]$ and $ipids6[17]$. If both time points show a $K+1$ increment (case1 in Figure~\ref{fig:example2}), we can determine that the NAT-penetration hole does not exist. If we do not observe an increment in both time points (case2 in Figure~\ref{fig:example2}), the NAT-penetration hole exists. The main point is that the retransmission occurs just in time to increase the IPID between the two measurements. In other situations, we cannot draw a confident conclusion.

However, the IPID increment does not always occur at the right time point. They may occur before or after the time point we get in $R$. The worse case is that the ``$K+1$'' IPID increment in one retransmission is divided into two different time points, before and after the timer being recorded.
Fortunately, these retransmission jitters
are always occurring in two nearby time points. 
This gives us the opportunity to simply extend the sampling point to its nearby time point to address these jitters. For example, the IPID increment between $ipids6[6]$ - $ipids6[5]$ is extended to $ipids6[6]$ - $ipids6[4]$ or $ipids6[7]$ - $ipids6[5]$, which can help us to make a more robust decision.

%% file: 5_measurement.tex
\section{Implementation}\label{sec:impl}
Our prototype system consists of three key components: the \textit{IP filter}, the \textit{outpost identifier}, and the \textit{penetration detector}. Those components are used in multiple places in the system. Their implementations are described as follows.

\vspace{3pt}\noindent$\bullet$
The IP filter's implementation is  based on ZMap~\cite{durumeric2013zmap}. Our Internet-wide scanning targets port 80, because it is the most common port used on the Internet.  
Our scanning also follows the ethics policy as outlined by ZMap, such as adding a blacklist and declaring the scanning purpose.

\vspace{3pt}\noindent$\bullet$
The outpost identifier is implemented with 400 lines of Python code. 
To expedite our scanning, we implement the outpost identifier by using a coroutine-based Python networking library (Gevent~\cite{gevent}), which provides a high-level synchronous API for network programming.
The IPID probe module $ipids0$ sets parameters $N$=10, $T$=1 as in the previous work~\cite{pearce2017augur}. The IPID probe module ($ipids1$, $ipids2$, and $ipids3$) sets parameters $N$=4, $T$=1.
For most operating systems, only the first SYN packet in the duplicated SYNs\footnote{The duplicated SYNs have the same source/destination port and source/destination IP address, but a different sequence number that is randomly generated.} is retransmitted.  
The spoofed packets used in the outpost identifier are the duplicated SYNs.
The selection is to ensure that only the first spoofed packet is retransmitted to minimize the network background traffic we induced.

\vspace{3pt}\noindent$\bullet$
The penetration detector is implemented with 450 lines of Python code. 
To speed up this process, we implement the penetration detector by using the traditional Python multi-threading package.
In order to induce multiple retransmissions,
the spoofed packets used in the penetration detector are not duplicated SYNs. Instead, each one has a different sequence number (SEQ), port, and IPID value in the TCP/IP packet header.
In addition, our scanner uses the \textit{iptables}~\cite{iptables} to enable the firewall rule below to drop the RST packets.
\begin{lstlisting}[language=sh, backgroundcolor=\color{white}]
iptables -A OUTPUT -s 46.XX.XX.35 -p tcp --dport 80 --tcp-flags RST RST -j DROP
\end{lstlisting}
This rule monitors the output chain and drops any RST packet whose source IP address is equal to 46.XX.XX.35.
In this case, the RST packet trying to reset the SYN-ACK packet is dropped, so the outpost continues to retransmit the SYN-ACK packet until the timeout is reached.

\begin{figure}
\centering
\includegraphics[width = 0.85\linewidth]{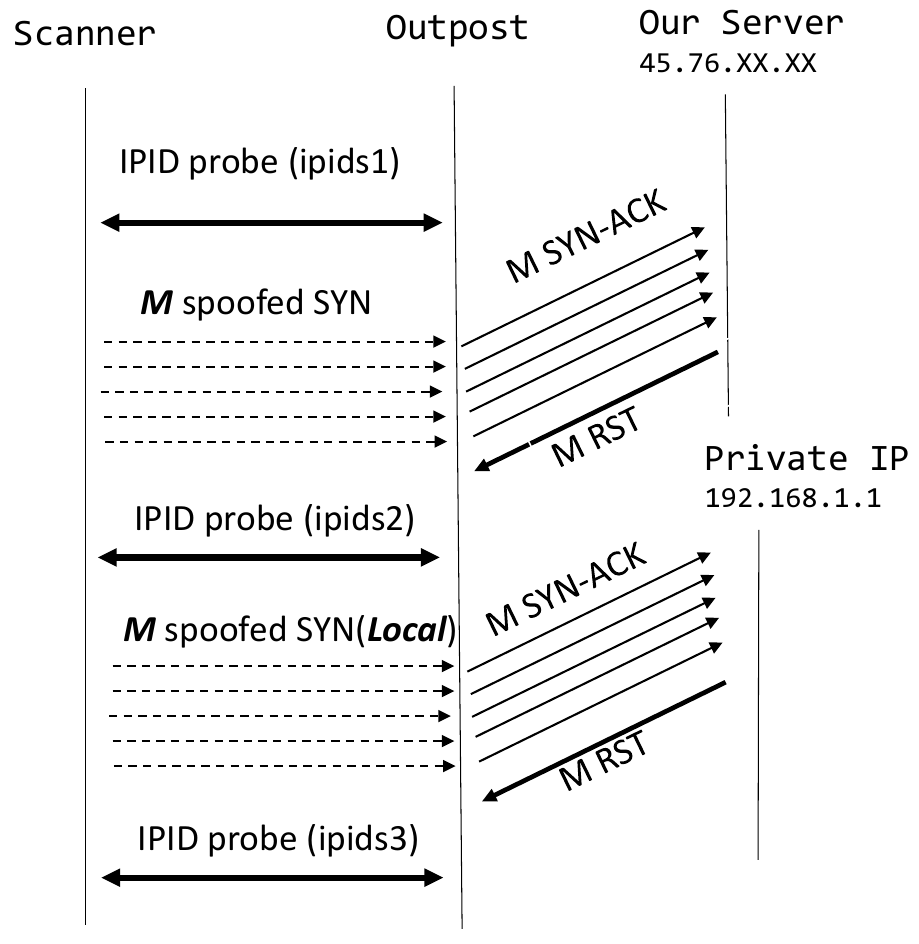}
\vspace{-10pt}
\caption{The outpost selection.}
\label{fig:outpost}
\end{figure}

\vspace{3pt}\noindent\textbf{ Settings.} \label{subsec:scanner_requirement}
The network where our scanner deploys should have the following settings.
These settings are about the environment of the scanner rather than the victim networks. We have found many such networks, allowing us to conduct the scanning described in the next section.

\vspace{3pt}\noindent$\bullet$\textit{ Accepting non-SYN TCP.}
A traditional stateful firewall~\cite{gouda2005model} rejects any SYN-ACK packets (in the output chain) when the first packet in a TCP session is not a SYN packet. This operation is used to avoid  SYN-ACK flood attacks. Most of the virtual private server (VPS) providers (such as Amazon Web Services and Microsoft Azure) strictly follow this policy.  
There are still many VPS providers (such as Vultr~\cite{vultr}) allowing users to disable the firewall, which makes it possible for doing SYN-ACK probes.
    
\vspace{3pt}\noindent$\bullet$\textit{ No egress filtering.} 
IP spoofing is often used to invoke DoS attacks against a target device or the surrounding infrastructure. 
To prevent attackers from launching an outbound malicious attack using IP spoofing, the VPS providers enable the egress filtering in the output chain to block any packets whose source IP addresses do not belong to their respective networks.
In fact, there are many little-known VPS providers that allow virtual machines
with IP spoofing enabled.

\section{Evaluation and Measurement}\label{sec:measurement}
In this section, we show the experiments that validate our method of revealing the existence of NAT-penetration hole via side channel. We then report our Internet-wide scanning results of the NAT-penetration test. 

\subsection{Evaluation }\label{subsec:validation}

We validated the effectiveness of our method in a controlled network and its robustness in a real network.

\vspace{3pt}\noindent\textbf{Testing in a controlled network.} We conducted our controlled experiment on a home network environment.
The network topology of the home network was similar to Figure~\ref{fig:attack-model}. 
There were 6 devices in the home network, including one Mi home security camera (continuously monitoring), one Mi Box (playing TV series), one Samsung galaxy TAB S7, one home server, and two smartphones. 
These devices had a same IP address prefix ``192.168.2.'' dynamically assigned by router A using DHCP~\cite{dhcp}. 
Router A was connected to the Internet and was assigned an IP address by another router B.
The IP address of router B was ``XX.XX.0.1'' and the outside IP address of router A was ``XX.XX.0.100''.  Our scanner was connected to router B and assigned an IP address (``XX.XX.0.101'').

In the advanced NAT setting, an HTTP server (port 8081) running on the home server was forwarded to router A (XX.XX.0.100:8081). This is a common setting\footnote{Port forwarding is used in the opposite direction compared to NAT. For example, a policy can be set on the router to forward ``any traffic originating in the Internet destined to the router's public interface with a port number of 80'' to the IP address (192.168.1.1) of your web server.
} when users want to access their home servers inside the gateway. 
Router A was an ideal outpost that bridged the local network and the public network.
We used it to identify the existence of NAT-penetration holes and further to reveal the private IP addresses behind router A (by testing all IPs in the prefix). 
When the above devices in the home network were all alive, we can successfully identify them (i.e., the NAT-penetration hole exits, which allows spoofed packet reaching the local network). 
Even if the devices were playing videos, the experiment received the same results.
We randomly turned off some devices and repeated the experiment. We were always able to identify the running devices correctly.

To show the threat and its consequence, we included laptop C (192.168.2.108) (running 14.04 with Linux kernel 4.4) into the network and forwarded
its SSH requests to router A (XX.XX.0.100:22). There was an SSH connection between  laptop C and home server D (192.168.2.101).
The IP addresses of laptop C and home server D had already been revealed.
Inspired by Cao et al.'s work~\cite{cao2016off}, we launched an off-path TCP attack to reset the SSH connection between C and D. 
By exploiting the challenge-ACK side channel, 
our scanner outside router A can successfully infer the existence of the TCP connection between devices C and D. 
Besides, we also inferred the four-tuple (source IP, source port, destination IP, and destination port) and the sequence number used in the TCP connection.
The off-path TCP connection reset attack was successfully launched
based on these steps.

\vspace{5pt}\noindent\textbf{Testing in a real network.}
We also validated our method in a real network to reveal the private IP addresses in the local network. 
The network was not owned by us. What we knew about this network is described as follows.
The publicly accessible outpost was a LANCOM 1781VAW router.
An IoT device (Solar-Log 1200) in the local network exposed its HTTP interface with the outpost's IP address by port forwarding. 
The local configuration of the IoT device was publicly accessible on the HTTP interface.
The private IP address of the IoT device was ``192.168.178.23'' and the IP address of the local gateway was ``192.168.178.1''.
We communicated with the outpost to test whether we could reveal the existence of a NAT-penetration hole and identify the existence of the two private IP addresses (``192.168.178.1'' and ``192.168.178.23'') in the ``192.168.178.1/16'' subnet. This testing was repeated 10 times.
We were able to successfully identify the existence of these two private IP addresses in all the tests. This suggests that the NAT-penetration hole stably exists on the target network.

Interestingly, the first three testings (among the above 10 times) could identify another three IP addresses (besides the two we already know) in the subnet. We further investigated to understand the reason. The first three tests were conducted during the working hours, but the other tests were around the midnight.
We believe that the devices corresponding to the three IP addresses were on and off during different hours of the day.
To support our hypothesis, we further monitored these five private IP addresses for 24 hours with a one-hour interval (i.e., evenly testing 24 times within a day).
We observed that the five IP addresses are all alive during the working hours, but only two IP addresses remained alive during the entire 24 hours.
In addition, the two devices that always stayed alive were either the gateway or some IoT devices.
This observation is not surprising, 
as personal devices are on and off as needed, but the IoT devices and the gateway are always alive. However, the exposure of the detailed on/off information may lead to privacy leakage. It allows  attackers to know the exact time when the victim user is at home and related personal activities, as well as the functions/types of the personal devices running on the private IP addresses.

\begin{table}
\footnotesize
\centering
\caption{The setup of our scanner.}
\label{table:machines}
\begin{tabular}{c c c c}
\toprule
    Stage & VPS & \ Machines  & Time Cost \\ \toprule
    Pre-Filtering & Vultr & 5  & 5h \\
    Outpost selection & Justhost.ru & 4  & 4 days \\ 

    NAT-penetration detection & Justhost.ru & -- & \ -- \\

    \toprule
\end{tabular}
\vspace{-10pt}
\end{table}

\begin{table*}
\footnotesize
\centering
\caption{\hlt{Internet-scale measurements.  
\hlta{
}
}}
\vspace{-2pt}
\label{table:real_world_attacks}
\begin{tabular}{L{1.6cm} C{3.5cm} C{3cm} C{1.3cm} C{1.2cm} C{1.2cm} C{1.4cm} C{1.2cm} }
\toprule
\multicolumn{1}{L{1.6cm}}{\textbf{Date}} & \multicolumn{1}{C{3.5cm}}{\textbf{Response to SYNACK ($IPID \neq 0$)}} & \multicolumn{1}{C{3cm}}{\textbf{Response to SYN ($IPID \neq 0$)}} & \multicolumn{1}{C{1.3cm}}{\textbf{Shared IPID}} & \multicolumn{1}{C{1.2cm}}{\textbf{Ideal IPID}} & \multicolumn{1}{C{1.2cm}}{\textbf{Spoofed IP}} & \multicolumn{1}{C{1.4cm}}{\textbf{Spoofed private IP}} & \multicolumn{1}{C{1.2cm}}{\textbf{Outpost}} \\ \toprule

2020/09/23 & 127,994,378 (62,826,277) & 12,645,088 (4,536,034) & 1,807,117 & 726,510 & 318,165 & 243,828 & 30,691  \\
2020/09/29 & 126,696,816 (62,051,373) & 12,576,566 (4,487,145) & 1,872,570 & 735,449 & - & - & - \\ 
2020/10/03 & 127,175,516 (62,404,431) & 12,531,861 (4,470,520) & 1,985,381 & 833,899 & 237,578 & 219,920 & 28,711  \\
2020/10/09 & 125,103,745 (61,222,297) & 12,646,871 (4,501,887) & 1,964,726 & 824,029 & 361,820 & 255,448 & 30,992  \\  \toprule
2020/10/12 &  47,512,365 (25,990,416) & 6,815,132 (2,490,381) & 1,481,428 & 637,025 & - & - & -   \\ 
2020/10/16 &  52,365,825 (28,563,207) & 7,478,645 (2,711,043) & 1,196,792 & 487,781 & 220,005 & 151,049 & 20,522  \\ 
2020/10/19 & 52,679,788 (28,753,731) & 7,556,356 (2,715,273) & 1,202,162 & 478,975 & 223,326 & 137,335 & 18,840 \\
\toprule
\end{tabular}
\vspace{-10pt}
\end{table*}

\subsection{Internet-wide Scanning}\label{subsec:world-setup}

We conducted several Internet-wide scans to systematically understand the NAT-penetration issues. The experiment setup and the summary of our scanning are shown below.

\vspace{5pt}\noindent\textbf{Experiment setup.} 
For generality, we deployed our scanner on cloud platforms. 
Table~\ref{table:machines} lists the detailed setup information of the scanner.
The outpost selection and the NAT-penetration detection were deployed on the cloud service provider Justhost~\cite{justhost_ru}. The pre-filtering was running on another cloud service provider (Vultr~\cite{vultr}).
Our scanner should satisfy two conditions (``Accepting non-SYN TCP'' and ``No  egress filtering'') as mentioned in Section~\ref{subsec:scanner_requirement}. Justhost satisfies both conditions, but it limits the transmission rate of SYN-ACK packets. 
This rate limit slows down the pre-filtering that
 needs to quickly send packets to test the entire IPv4 address space. 
Fortunately, the pre-filter only needs to send SYN-ACK packets without spoofing source IP addresses. Therefore, we ran the pre-filter on another cloud service provider Vultr~\cite{vultr}, in which it can send SYN-ACK packets rapidly with a high bandwidth.

The pre-filter worked on 5 Vultr machines (1 CPU, 1 GB memory, 1 Gbps bandwidth).
At a speed of 50K packets per second, it took the pre-filter about 5 hours to finish the Internet-wide scanning. 
The outpost selection worked on 4 Justhost machines (2 CPU, 512 Mb memory, 200 Mbps bandwidth). It took the outpost selection about 4 days to finish the Internet-wide scanning. 
To reduce the background traffic on the Internet, the outpost selection only considered the outposts with the IPID noise below 6 per second.
Note that the total time of our adaptive scanner was not affected by the number of spoofed packets.

\vspace{5pt}\noindent\textbf{Summary of the scanning.} 
In a period of five weeks, we conducted seven Internet-wide experiments. Table~\ref{table:real_world_attacks} presents the results in detail.
``Shared IPID'' is the number of hosts having a shared IPID counter.
{\color{black} Note that this number reported here is our scanning result 
(after two pre-filters). It is not the whole hosts with a shared IPID on the IPv4 space. For the number of those hosts having a shared IPID on the whole IPv4 space, we have separately validated, which is consistent with 22 million in ~\cite{pearce2017augur}.}
``Ideal IPID'' is the number of hosts having a shared IPID counter with an IPID noise less than 6.  ``Spoofed IP'' is the number of hosts blocking spoofed public IP addresses. ``Spoofed private IP'' is the number of hosts blocking spoofed private IP addresses.
The first four experiments targeted the entire IPv4 address space.
To speed up the scanning, for the last three experiments, we only scanned the /16 network blocks that had at least one outpost in the previous five experiments.
The pre-filtering found about 4.5 million reachable candidate hosts. Approximately 2 million of these hosts were found to use a shared, monotonically increasing IPID.  
In each of the eight experiments, our scanner identified about 30,000 outposts, geographically distributed across 154 countries and 6,374 organizations in the world.

Once the vulnerable outposts are identified, we can reliably reveal the existence of NAT-penetration holes in the real networks, as shown in Section~\ref{subsec:validation}. 
However, due to the ethical consideration, we did not conduct the real NAT-penetration test on the outposts in the Internet-wide scanning, but it is a straightforward task that can be performed by administrators of the vulnerable networks.

\subsection{Analysis and Understanding} 
Our analysis is focused on the outposts, as their existence directly leads to high possibility of NAT-penetration in the vulnerable networks as mentioned above. 

\begin{figure}
\vspace{-6pt}
\centering
\includegraphics[width = 0.9\linewidth]{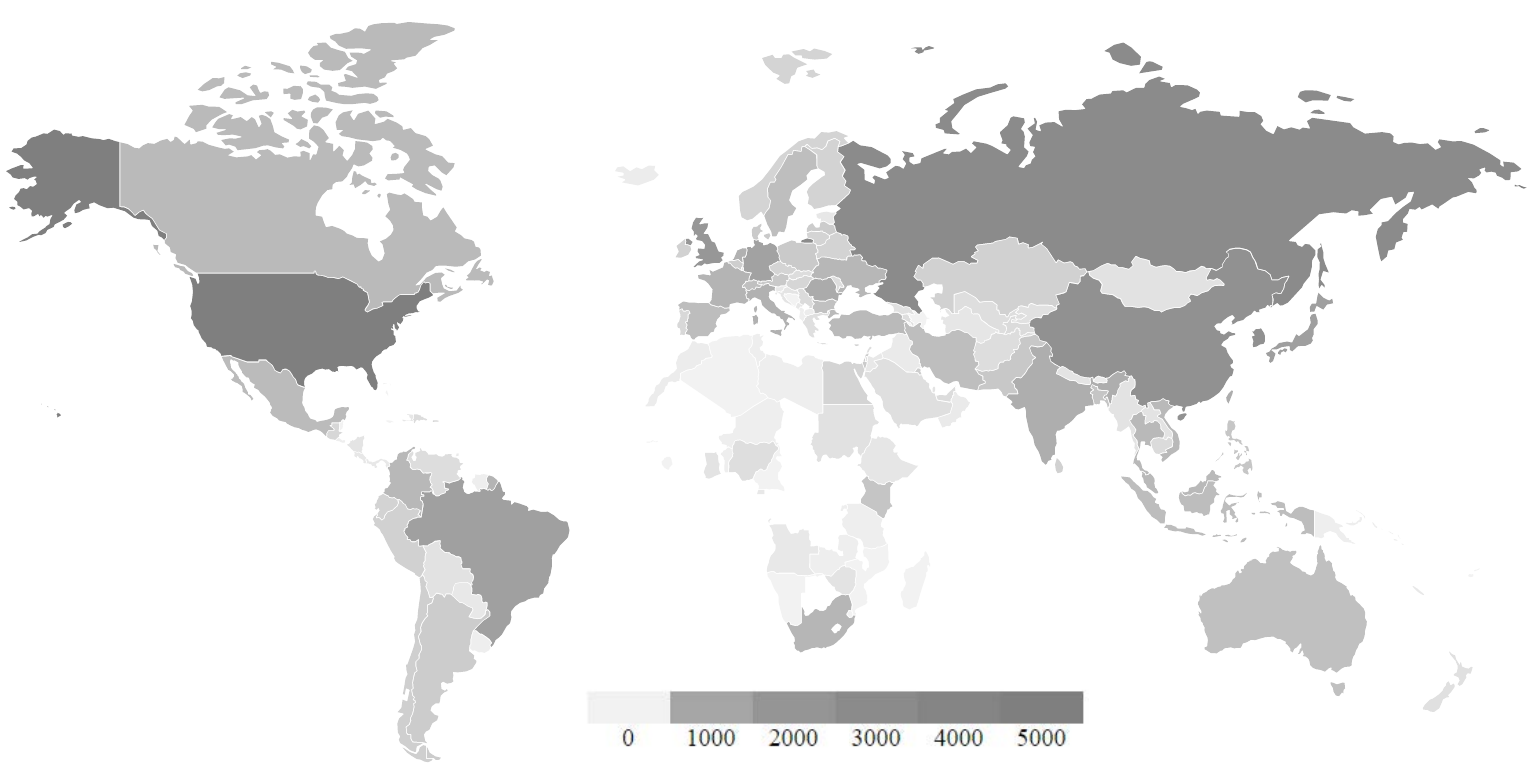}
\caption{Global distribution of the outposts.}
\label{fig:world-map}
\vspace{-10pt}
\end{figure}

\vspace{5pt}\noindent\textbf{Distributions.} 
Maxmind's GeoIP2~\cite{geoip} database is a commercial IP geolocation database, which provides the mapping between an IP address and its geolocation information. We used this database to retrieve the geolocation information (country, city, ISP, and autonomous system number) of the outposts.
Figure~\ref{fig:world-map} shows the geographical distribution of the outposts 
over the world, as determined by their geolocations.
The number of the outposts in each country is ranked and illustrated with various shades of darkness in the figure. The darker color indicates more outposts in the country. As the map shows, USA, Russian, Korea, and China are the top-4 countries in terms of the number of outposts. 
We found that the distribution of outposts among countries follows a long-tail: more than 65\% of the total outposts are from the top-10 countries.
These outposts are distributed in 4,164 different ISPs and 4,364 autonomous systems.
Table~\ref{table:isp_distribution} lists the top-10 ISPs that own the most outposts. They are all well-known ISPs in different countries. 
There are 2,020 (49\% among 4,164) ISPs having at least 2 outposts. The outposts in these ISPs occupy about 93\% of the total outposts. 

\vspace{5pt}\noindent\textbf{IP dynamics.}
We also analyzed the outposts' IP dynamics. Table~\ref{table:org_dynamic} lists the details. ``ISP'' is the Internet service provider of the corresponding IP addresses. The outposts we collected in Section~\ref{subsec:world-setup} are clustered by different network blocks (/16, /20, /24) and ISPs. ``U1'' is the set containing all elements that appear both on Sep 23, 2020 and Oct 03, 2020. ``U2'' is the set containing all elements that appear both on Oct 03, 2020 and Oct 09, 2020. 
The total number of the outposts seems stable across the experiments, but the actual IP addresses are substantially different. Only 23\% of the IP addresses appear in all three experiments. The IP addresses of the outposts are dynamically changed.
The ISP level shows lower dynamics than the IP level.
As shown in Table~\ref{table:org_dynamic}, about 70\% of the ISPs appear in two experiments and about 60\% of the ISPs appear in all three experiments.

\begin{table}
\footnotesize
\centering
\caption{List of top 10 ISPs}
\label{table:isp_distribution}
\vspace{-5pt}
\begin{tabular}{L{0.4cm} C{4.7cm} C{1cm}}
\toprule
 & ISPs &  \# \\  \toprule
1 & TalkTalk & 1,445  \\
2 & LG DACOM Corporation & 1,200  \\
3 & IKGUL & 1,193 \\
4 & Hangzhou Alibaba Advertising Co.,Ltd & 828 \\
5 & Deutsche Telekom AG & 646 \\
6 & Rostelecom & 587 \\
7 & Telekom Romania & 538 \\
8 & Hong Kong Broadband Network & 381 \\
9 & Korea Telecom & 365 \\
10 & China Unicom Liaoning & 336 \\
\toprule
\end{tabular}
\vspace{-10pt}
\end{table}

\vspace{5pt}\noindent\textbf{Device types of the outposts.}~\label{subsec:device_type}
We further investigated the device types of the outposts.
We randomly chose 100 outposts from one Internet-wide experiment.
We manually accessed the webpages hosted on the outposts' HTTP server (port 80) via a web browser. 
Using the device fingerprinting method~\cite{xuan2018acquisitional},
we attempted to give each outpost a device label, including the manufacturer, the device type, and the device model.

We successfully identified the device types for 83 outposts\footnote{There were 10 outposts not running any services on port 80, and 7 outposts that we cannot give a confident device label.}.
Among the 83 outposts, 47 devices were routers, switches, and other network devices.
The manufacturers of the outposts span across 14 brands. The top three are HUAWEI, Cisco, and Technicolor. 
Besides the networking equipment, 12 devices (13.8\%) installed Windows operating system, and ran a server program (such as Microsoft-IIS, Microsoft-WinCE, and Windows Management Framework, etc.) on port 80. 
17 devices (19.5\%) ran a Vmware ESXi hypervisor~\cite{esxi}. 
The virtual machines deployed on the hypervisors with the private IP addresses were identified as the local network devices.
Seven devices were IoT devices, such as Hikvision video surveillance equipment, Leica Geosystems GR30 GNSS reference server, OKI C9650 printer, Solar-Log 1200 energy management system, and Pl@ntVisor supervision and monitoring system.

We also analyzed the device types of the outposts in the same ISPs.
We randomly chose 55 outposts across the top 13 ISPs (see Table~\ref{table:isp_distribution}) and manually identified the device types of the outposts. Most of the devices (49 out of  55) were routing/switching devices.
We found that the outposts in the same ISPs were similar devices.
For example, ``TalkTalk'' used  Technicolor TG582n and Thomson TG789vn routers; ``Hong Kong Broadband Network'' used Ruijie RSR series routers.

\begin{table}
\footnotesize
\centering
\caption{The outposts clustered by subnets. 
}
\label{table:org_dynamic}
\begin{tabular}{L{1.3cm} C{0.8cm} C{1cm} C{1cm} C{1cm} C{0.8cm} }
\toprule
\textbf{Date} &	\textbf{IPs} &	\textbf{/24 subnets} & \textbf{/20 subnets} &  \textbf{/16 subnets} & \textbf{ISPs} \\ \toprule
2020/09/23 & 30,691 & 23,236 & 16,772 & 8,521 & 4,388 \\ 
2020/10/03 & 28,711  & 22,059 & 15,832 & 8,182 & 4,140 \\ 
2020/10/09 & 30,992 & 22,842 & 16,306 & 8,309 & 4,164 \\ 
\hline
U1 & 6,697 & 8,270 & 8,011 & 5,810 & 2,755 \\ 
U2 & 8,342 & 9,089 & 8,216 & 5,781 & 2,797 \\ 
Total & 76,447 &	50,244 & 31,674 &	12,734 & 6,734 \\
\toprule
\end{tabular}
\vspace{-10pt}
\end{table}

\vspace{5pt}\noindent\textbf{Misconfigured legacy systems.}
As mentioned in Section~\ref{subsec:outpost_requirement}, a shared IPID counter and no ingress filtering are the two important requirements for conducting the NAT-penetration test.
The no-ingress-filtering issue is a misconfiguration which exposes a ``tunnel'' between the public network and the local network. 
A shared IPID counter is one vulnerability of building TCP/IP side channels, which give remote attackers the opportunity to verify that this ``tunnel'' indeed exists.
We investigate how common the two individual problems are on the Internet.
The vulnerable shared IPID counter, which is a very old problem, has been discussed in  1998~\cite{ipid-original}. 
However, our study shows that this legacy problem still exists  (more than 2 million hosts we found on the Internet) more than 20 years later. 

For the no-ingress-filtering issue, as shown in Table~\ref{table:real_world_attacks}, there are 56.2\% of the hosts with the ideal global IPID (out of 0.824 million) having the problem. 
{\color{black} 
The number is consistent with the 2019 study~\cite{luckie2019network} that reports 59.2\% of ASes do not filter any inbound packets.
}
Tens of thousands of edge devices do not even filter any packets that have a spoofed private IP address. The filtering fails in both well-known ISPs and well-known device manufacturers. This is the main cause of the PIRSC vulnerability, allowing attackers outside gateway to penetrate NAT boxes.

\vspace{5pt}\noindent\textbf{Vulnerability notification.} Notifying all the organizations with discovered vulnerabilities is a time-consuming task. We have notified the top 10 organizations.
We found the organization's email contact from their official website or WHOIS information.
We notified them of the IP addresses that we found vulnerable. However, it is disappointing that no organization has yet replied to our notification. 
{\color{black}
We will continue working on the notification process and follow the guideline for responsible disclosure. 
}

%% file: 6_discussion.tex
\section{Discussions}\label{sec:discussion}

In this section, we discuss the limitations of our experiments, mitigation actions that should be taken, and the ethics.

\vspace{5pt}\noindent\textbf{Limitations}. 
Our scanner only targets the hosts with port 80 open. {\color{black} Also, to guarantee the politeness, we discard the hosts that satisfy the outpost requirements but have too much IPID noise. 
} 
These limitations reduce the number of the outposts that can be potentially identified on the Internet. 
{\color{black}
However, our primary goal is not to find all outposts on the Internet or NAT-penetration holes through the outposts. Instead, we attempt to reveal the wide existence of NAT-penetration holes and validate the effectiveness of our exploration method, as well as gain a deep understanding of the NAT penetration issue in an efficient and non-aggressive manner.
The number of vulnerabilities found in our study has already demonstrated the seriousness of the problem.}
If other researchers plan to conduct more comprehensive analysis for their own organizations based on our proposed approach, they can easily increase the number of the spoofed packets and target all open ports on the target hosts.

{\color{black}
We acknowledge that many modern systems do not use shared IPID, and IP spoofed packets can be blocked if ISPs perform ingress filtering (i.e., the proposed penetration can be easily throttled by using current defenses).
However, our real-world measurement study demonstrates that some legacy problems still widely exist today, even though they have been solvable for a long time.  
For example, the study~\cite{pearce2017augur} conducted in 2017, the study conducted in 2018~\cite{salutari2018closer}, and our study all show that there are still a significant number of hosts (22.7 million, which is about 16\% among 140 million reachable hosts.) with shared IPIDs on the Internet.
}

\vspace{5pt}\noindent\textbf{Mitigation.}
Based on the results of our measurement study, we have identified several mitigation strategies. We observe more than 2 million hosts with a shared IPID counter, which is an old vulnerability dated back to 1998. We also observe a large number of ISPs and manufacturers with unsafe firewall configurations (no ingress filtering and not blocking spoofed private IP addresses). Obviously, the mitigation of this weakness needs collaborations among administrators, manufacturers, and ISP owners.

First, administrators should regularly check their devices and update the software/firmware when security patches are available. Ensuring software/firmware up-to-date is the most effective mitigation to prevent network devices from being attacked. For the systems that are hardly to be updated in a timely fashion (such as industrial control system), administrators should deploy specific firewalls.
Second, device manufacturers should increase their efforts to avoid misconfigurations, and notify their customers about updating their devices more effectively (e.g., using an automatic update mechanism).
Finally, ISP owners should ensure the network-level security for their customers. In fact, we observe that a large number of ISP owners do not deploy the basic filtering for the ingress traffic. 
We suggest that ISP owners enable the following settings in the network-layer firewalls: (1) rejecting non-SYN TCP packets. If the first packet is not an SYN packet, this packet should be dropped. This setting can be enabled in most of the commercial firewalls.
(2) preventing IP spoofing. For end-users, detecting IP spoofing is virtually impossible. The task must be conducted by ISPs. Many methods are practical, such as blocking special IP addresses (e.g., the private IP addresses), access control lists (ACLs), and reverse path forwarding (RPF).

\vspace{3pt}\noindent\textbf{Ethics.} 
As mentioned earlier, we follow the ethical policy documented by ZMap~\cite{durumeric2013zmap} to conduct the Internet-wide scanning (pre-filtering). 
We also limit the packet transmission rate at 0.6 packet per second to each host, which is even less than that of the previous scanning method based on the TCP backlog side channel~\cite{zhang2015original}.
We try our best to minimize the resource consumption at the target machine by using  SYN-ACK packets, instead of SYN packets (in the IPID probe module). 
In addition, during the stage of monitoring the target's IPID increment, we only record the IPID field in the TCP packet header. We refrain ourselves from gathering any information not supposed to collect, such as identity-related data.

In addition, our scanning only identifies candidate outposts that we believe to be vulnerable, and no actual penetration is performed for ethical considerations because these networks are not owned by us. Instead, we conducted experiments in our own networks to confirm that the observed vulnerability patterns indeed lead to a high probability of successful penetrations.

%% file: 7_relatedwork.tex
\section{Related Work}
\label{sec:relatedwork}

\noindent\textbf{Finding network devices behind NAT.}
In previous network measurement studies, their focuses are on publicly accessible networks~\cite{antonakakis2017understanding,Cai2010,cui2010quantitative,durumeric2015search,durumeric2014matter,durumeric2013zmap,fachkha2017internet,feng2016characterizing,heidemann2008census, leonard2010demystifying, Hershel2014}.
However, along with the pervasiveness of Internet-of-Things (IoT) systems,  more devices have been deployed in the home networks.
Researchers have shown the feasibility of detecting home devices behind NAT~\cite{acar2018web,kumar2019all,rytilahti2020using}.

Acar et al.~\cite{acar2018web} proposed to detect the presence of IoT devices in the local network
by exploiting malicious web pages or third-party ads. The malicious scripts could bypass the same-origin policy by exploiting error messages on the HTML5 MediaError interface or by DNS rebinding attacks. This attack allows an adversary to extract private information (e.g., serial numbers, user names, or geolocations) from the local devices or even control these devices (e.g., rebooting or playing arbitrary videos).
Kumar et al.~\cite{kumar2019all} partnered with Avast Software, a popular anti-virus company, whose security software enables its customers scan their local networks for IoT devices that support weak authentication or have remotely exploitable vulnerabilities. Leveraging the data collected from user-initiated network scans in 16 million households that had agreed to share
data for research and development purposes, they provided large-scale empirical analysis of IoT devices in the home networks.
Rytilahti et al.~\cite{rytilahti2020using} took advantage of some application-layer middlebox protocols (e.g., UPnP, NAT-PMP/PCP, and network proxies), which enabled end-to-end connectivity to those ``hidden'' devices, to peek network devices behind NAT gateways.
The above methods either require the cooperation of  end-users or malicious scripts coincidentally triggered by the user's browser, which limits the generality of their approaches.

\vspace{6pt}\noindent\textbf{TCP/IP side channel for network measurement}.
{\color{black}
The TCP/IP side channel has been investigated in many prior studies, such as the TCP/IP side channel vulnerabilities and attacks~\cite{cao2016off}\cite{feng2020off}\cite{klein2019ip}, measurement studies on TCP/IP side channel on the real-world~\cite{salutari2018closer}~\cite{west2006tcp}\cite{quach2017investigation}.
Our work is a measurement study that leverages the shared IPID side channel to conduct NAT penetration measurement. Therefore, we mainly list the prior studies that exploit TCP/IP side channels to conduct Internet measurements where
conventional methods cannot achieve.
}

The first work of using the TCP/IP side channel for network scanning was dated back to 1998. 
The researcher~\cite{ipid-original} released a new TCP scan technique (later called idle scan~\cite{idle}) in a mailing list. 
This technique enables the scanner to scan a target machine without using the scanner's own return IP address in a packet sent to the target. Instead, the scanner sends spoofed packets with a zombie's IP address to the target machine. The zombie machine has a shared IPID counter. By observing the IPID change in the zombie, we can infer whether a port on the target is closed or open. The idle scan was implemented in Nmap~\cite{nmap}.

Later, Ensafi et al.~\cite{ensafi2010idle} proposed a scan method that does not need to send any packets (not even spoofed packets) to the target machine at all. It is the first conceptual approach that can infer the existence of local devices behind NAT/firewall middleboxes. However, this method is required to fill the TCP backlog, causing a DoS attack on zombies. 
To avoid DoS attacks at zombies, Zhang et al.~\cite{zhang2015original} proposed to fill half of the backlog (affected the Linux kernel versions 2.3 and later). However, such a method is required to send at a rate of 5 packets per second for about 60 seconds.  In addition, different systems have different TCP backlog sizes. The process of inferring the backlog size is also time-consuming. 
For these reasons, such a method is incompetent for large-scale network scanning.

Alexander et al.~\cite{alexander2015off} leveraged the TCP backlog side channel to measure off-path round trip times between arbitrary Internet end hosts.
Ensafi et al.~\cite{ensafi2014detecting} proposed to discover the packet drops (e.g., due to censorship) between two arbitrary IP addresses on the Internet based on the shared-IPID side channel. 
 Augur~\cite{pearce2017augur} utilized the shared-IPID side channel to measure the reachability between two Internet locations without directly controlling a measurement vantage point at either location.
{\color{black} The most recent work of using the shared-IPID side channel is SMap~\cite{dai2020smap}, which is used to conduct IP spoofing measurements on the Internet. The method used in the spoofability checker of our work is similar to this work.} 
Note that the previous studies above all focus on the publicly accessible IP addresses.
By contrast, our work focuses on private IP addresses behind NAT.
We use the TCP/IP side channel to infer the reachability of those private local networks behind NAT, which has not yet been fully investigated before.

%% file: 8_conclusion.tex
\section{Conclusion}
\label{sec:conclusion}

We conducted an Internet-wide study on the NAT-penetration problem by leveraging TCP/IP side channels.
Based on the investigation of several TCP/IP side channels documented before, we found that the shared-IPID side channel is the most suitable candidate for performing the Internet-wide penetration test. We considered generality, ethics, and robustness for the development of our scanning technique. Specifically, our scanner can automatically adjust its strategies and make robust decisions for different network scenarios. Our scanner is able to complete the whole IPv4 address space scanning in five days, even if it sends packets at a very non-aggressive rate.  
The evaluation results show that our scanner is effective and efficient in both the controlled network and the real network scenarios.
Our measurement results reveal that the NAT penetration problem widely exists on the Internet across 154 countries and 4,146 different organizations. We call for serious attention and efforts from the network security community to fully address this Internet security threat caused by legacy problems.

%% file: main.bbl
\begin{thebibliography}{10}

\bibitem{ipid-original}
Bugtraq mailing list archives: new tcp scan method.
\newblock \url{ https://seclists.org/bugtraq/1998/Dec/79}.

\bibitem{gevent}
A coroutine -based python networking library.
\newblock \url{http://www.gevent.org/}.

\bibitem{iptables}
{iptables - administration tool for IPv4 packet filtering and NAT}.
\newblock \url{https://linux.die.net/man/8/iptables}.

\bibitem{justhost_ru}
Justhost.ru.
\newblock \url{https://justhost.ru/}.

\bibitem{masscan}
{Masscan, Network Scanner tool for scanning Internet port}.
\newblock \url{https://github.com/robertdavidgraham/masscan}.

\bibitem{bcp38}
Network ingress filtering: Defeating denial of service attacks which employ ip
  source address spoofing.
\newblock \url{https://tools.ietf.org/html/bcp38}.

\bibitem{nmap}
Nmap, network security scanner tool.
\newblock \url{https://nmap.org/}.

\bibitem{vultr}
Ssd vps servers, cloud servers and cloud hosting by vultr.
\newblock \url{ https://www.vultr.com}.

\bibitem{idle}
Tcp idle scan.
\newblock \url{ https://nmap.org/book/idlescan.html}.

\bibitem{esxi}
Vmware esxi: The purpose-built bare metal hypervisor.
\newblock \url{https://en.wikipedia.org/wiki/VMware_ESXi}.

\bibitem{acar2018web}
{\sc Acar, G., Huang, D.~Y., Li, F., Narayanan, A., and Feamster, N.}
\newblock Web-based attacks to discover and control local iot devices.
\newblock In {\em Proceedings of the 2018 Workshop on IoT Security and
  Privacy\/} (2018), pp.~29--35.

\bibitem{alexander2015off}
{\sc Alexander, G., and Crandall, J.~R.}
\newblock Off-path round trip time measurement via tcp/ip side channels.
\newblock In {\em Proceedings of the 2015 IEEE International Conference on
  Computer Communications (INFOCOM'15)\/} (2015), IEEE, pp.~1589--1597.

\bibitem{antonakakis2017understanding}
{\sc Antonakakis, M., April, T., Bailey, M., Bernhard, M., Bursztein, E.,
  Cochran, J., Durumeric, Z., Halderman, J.~A., Invernizzi, L., Kallitsis, M.,
  Kumar, D., Lever, C., Ma, Z., Mason, J., Menscher, D., Seaman, C., Sullivan,
  N., Thomas, K., and Zhou, Y.}
\newblock Understanding the mirai botnet.
\newblock In {\em Proceedings of the {USENIX} Security Symposium (SEC'17)\/}
  (2017), pp.~1093--1110.

\bibitem{Cai2010}
{\sc Cai, X., and Heidemann, J.}
\newblock Understanding block-level address usage in the visible internet.
\newblock In {\em Proceedings of the ACM SIGCOMM Conference ({SIGCOMM '10})\/}
  (New York, NY, USA, 2010), ACM, pp.~99--110.

\bibitem{cao2016off}
{\sc Cao, Y., Qian, Z., Wang, Z., Dao, T., Krishnamurthy, S.~V., and Marvel,
  L.~M.}
\newblock Off-path tcp exploits: Global rate limit considered dangerous.
\newblock In {\em Proceedings of the 25th USENIX Security Symposium (SEC'16)\/}
  (2016), pp.~209--225.

\bibitem{chen2018off}
{\sc Chen, W., and Qian, Z.}
\newblock Off-path $\{$TCP$\}$ exploit: How wireless routers can jeopardize
  your secrets.
\newblock In {\em 27th $\{$USENIX$\}$ Security Symposium ($\{$USENIX$\}$
  Security 18)\/} (2018), pp.~1581--1598.

\bibitem{cui2010quantitative}
{\sc Cui, A., and Stolfo, S.~J.}
\newblock A quantitative analysis of the insecurity of embedded network
  devices: results of a wide-area scan.
\newblock In {\em Proceedings of the 26th Annual Computer Security Applications
  Conference ({ACSAC'10})\/} (Austin, Texas, USA, 2010), pp.~97--106.

\bibitem{dai2020smap}
{\sc Dai, T., and Shulman, H.}
\newblock Smap: Internet-wide scanning for ingress filtering.
\newblock {\em arXiv preprint arXiv:2003.05813\/} (2020).

\bibitem{dhcp}
{\sc Droms, R.}
\newblock Dynamic host configuration protocol. ietf rfc-2131, march 1997.,
  1997.

\bibitem{durumeric2015search}
{\sc Durumeric, Z., Adrian, D., Mirian, A., Bailey, M., and Halderman, J.~A.}
\newblock A search engine backed by internet-wide scanning.
\newblock In {\em Proceedings of the 22nd ACM SIGSAC Conference on Computer and
  Communications Security (CCS'15)\/} (2015), ACM, pp.~542--553.

\bibitem{durumeric2014matter}
{\sc Durumeric, Z., Kasten, J., Adrian, D., Halderman, J.~A., Bailey, M., Li,
  F., Weaver, N., Amann, J., Beekman, J., Payer, M., and Paxson, V.}
\newblock The matter of heartbleed.
\newblock In {\em Proceedings of the 2014 Internet Measurement Conference
  (IMC'14), Vancouver, BC, Canada, November 5-7, 2014}, pp.~475--488.

\bibitem{durumeric2013zmap}
{\sc Durumeric, Z., Wustrow, E., and Halderman, J.~A.}
\newblock Zmap: Fast internet-wide scanning and its security applications.
\newblock In {\em Proceedings of the 22th {USENIX} Security Symposium
  (SEC'13)\/} (2013), pp.~605--620.

\bibitem{ensafi2014detecting}
{\sc Ensafi, R., Knockel, J., Alexander, G., and Crandall, J.~R.}
\newblock Detecting intentional packet drops on the internet via tcp/ip side
  channels.
\newblock In {\em Proceedings of the International Conference on Passive and
  Active Network Measurement (PAM'14)\/} (2014), Springer, pp.~109--118.

\bibitem{ensafi2010idle}
{\sc Ensafi, R., Park, J.~C., Kapur, D., and Crandall, J.~R.}
\newblock Idle port scanning and non-interference analysis of network protocol
  stacks using model checking.
\newblock In {\em Proceedings of the 2010 USENIX Security Symposium (SEC'10)\/}
  (2010), pp.~257--272.

\bibitem{fachkha2017internet}
{\sc Fachkha, C., Bou-Harb, E., Keliris, A., Memon, N., and Ahamad, M.}
\newblock Internet-scale probing of cps: Inference, characterization and
  orchestration analysis.
\newblock In {\em Proceedings of Network and Distributed System Security
  Symposium (NDSS'17)\/} (2017), vol.~17.

\bibitem{feng2020off}
{\sc Feng, X., Fu, C., Li, Q., Sun, K., and Xu, K.}
\newblock Off-path tcp exploits of the mixed ipid assignment.
\newblock In {\em Proceedings of the 2020 ACM SIGSAC Conference on Computer and
  Communications Security\/} (2020), pp.~1323--1335.

\bibitem{feng2016characterizing}
{\sc Feng, X., Li, Q., Wang, H., and Sun, L.}
\newblock Characterizing industrial control system devices on the internet.
\newblock In {\em Proceedings of the 24th {IEEE} International Conference on
  Network Protocols (ICNP'16)\/} (2016), IEEE, pp.~1--10.

\bibitem{xuan2018acquisitional}
{\sc Feng, X., Li, Q., Wang, H., and Sun, L.}
\newblock Acquisitional rule-based engine for discovering internet-of-thing
  devices.
\newblock In {\em Proceedings of the {USENIX} Security Symposium (SEC'18)\/}
  (2018), pp.~327--341.

\bibitem{gouda2005model}
{\sc Gouda, M.~G., and Liu, A.~X.}
\newblock A model of stateful firewalls and its properties.
\newblock In {\em Proceedings of the 2005 International Conference on
  Dependable Systems and Networks (DSN'05)\/} (2005), IEEE, pp.~128--137.

\bibitem{heidemann2008census}
{\sc Heidemann, J.~S., Pryadkin, Y., Govindan, R., Papadopoulos, C., Bartlett,
  G., and Bannister, J.~A.}
\newblock Census and survey of the visible internet.
\newblock In {\em Proceedings of the {ACM} {SIGCOMM} Internet Measurement
  Conference (IMC'08)\/} (Vouliagmeni, Greece, 2008), pp.~169--182.

\bibitem{klein2019ip}
{\sc Klein, A., and Pinkas, B.}
\newblock From ip id to device id and kaslr bypass.
\newblock In {\em 28th USENIX Security Symposium (USENIX Security 19)\/}
  (2019), pp.~1063--1080.

\bibitem{kumar2019all}
{\sc Kumar, D., Shen, K., Case, B., Garg, D., Alperovich, G., Kuznetsov, D.,
  Gupta, R., and Durumeric, Z.}
\newblock All things considered: an analysis of iot devices on home networks.
\newblock In {\em Proceedings of the 28th USENIX Security Symposium (SEC'19)\/}
  (2019), pp.~1169--1185.

\bibitem{leonard2010demystifying}
{\sc Leonard, D., and Loguinov, D.}
\newblock Demystifying service discovery: implementing an internet-wide
  scanner.
\newblock In {\em Proceedings of the 10th {ACM} {SIGCOMM} Internet Measurement
  Conference (IMC'10)\/} (2010), pp.~109--122.

\bibitem{luckie2019network}
{\sc Luckie, M., Beverly, R., Koga, R., Keys, K., Kroll, J.~A., and Claffy, K.}
\newblock Network hygiene, incentives, and regulation: deployment of source
  address validation in the internet.
\newblock In {\em Proceedings of the 2019 ACM SIGSAC Conference on Computer and
  Communications Security (CCS'19)\/} (2019), pp.~465--480.

\bibitem{geoip}
{\sc MaxMind}.
\newblock Maxmind geoip2 database.
\newblock \url{https://www.maxmind.com/en/geoip2-services-and-databases}, 2021.

\bibitem{mi2019resident}
{\sc Mi, X., Feng, X., Liao, X., Liu, B., Wang, X., Qian, F., Li, Z., Alrwais,
  S., Sun, L., and Liu, Y.}
\newblock Resident evil: Understanding residential ip proxy as a dark service.
\newblock In {\em Proceedings of the 2019 IEEE symposium on security and
  privacy (SP'19)\/} (2019), IEEE, pp.~1185--1201.

\bibitem{pearce2017augur}
{\sc Pearce, P., Ensafi, R., Li, F., Feamster, N., and Paxson, V.}
\newblock Augur: Internet-wide detection of connectivity disruptions.
\newblock In {\em Proceedings of the IEEE Symposium on Security and Privacy
  (SP'17)\/} (2017), IEEE, pp.~427--443.

\bibitem{quach2017investigation}
{\sc Quach, A., Wang, Z., and Qian, Z.}
\newblock Investigation of the 2016 linux tcp stack vulnerability at scale.
\newblock {\em Proceedings of the ACM on Measurement and Analysis of Computing
  Systems 1}, 1 (2017), 1--19.

\bibitem{ramaiah2010rfc}
{\sc Ramaiah, A., Stewart, R., and Dalal, M.}
\newblock Rfc 5961: Improving tcp’s robustness to blind in-window attacks,
  2010.

\bibitem{rytilahti2020using}
{\sc Rytilahti, T., and Holz, T.}
\newblock On using application-layer middlebox protocols for peeking behind nat
  gateways.
\newblock In {\em Proceedings of the Network and Distributed System Security
  Symposium (NDSS'20)\/} (2020).

\bibitem{salutari2018closer}
{\sc Salutari, F., Cicalese, D., and Rossi, D.~J.}
\newblock A closer look at ip-id behavior in the wild.
\newblock In {\em International Conference on Passive and Active Network
  Measurement\/} (2018), Springer, pp.~243--254.

\bibitem{Hershel2014}
{\sc Shamsi, Z., Nandwani, A., Leonard, D., and Loguinov, D.}
\newblock Hershel: single-packet os fingerprinting.
\newblock In {\em Proceedings of the {ACM} {SIGMETRICS} / International
  Conference on Measurement and Modeling of Computer Systems (SIGMETRICS'14)\/}
  (2014), pp.~195--206.

\bibitem{west2006tcp}
{\sc West, M., and McCann, S.}
\newblock Tcp/ip field behavior.
\newblock Tech. rep., RFC 4413, March, 2006.

\bibitem{zhang2015original}
{\sc Zhang, X., Knockel, J., and Crandall, J.~R.}
\newblock Original syn: Finding machines hidden behind firewalls.
\newblock In {\em Proceedings of the 2015 IEEE International Conference on
  Computer Communications (INFOCOM'15)\/} (2015), IEEE, pp.~720--728.

\end{thebibliography}
